\documentclass[twocolumn,showpacs,preprintnumbers,amsmath,amssymb]{revtex4}
\usepackage{graphicx,braket}
\usepackage{dcolumn,comment}
\usepackage{bm}
\newcommand{\Tr}{ \mathrm{Tr}}
\begin{document}
\title{Lazy open quantum walks}

\author{Garreth Kemp$^{a}$}
 \email{gkemp@uj.ac.za}
\author{Ilya Sinayskiy$^{b,c}$ and Francesco Petruccione$^{b,c,d}$}
  \affiliation{$^{a}$Department of Mathematics and Applied Mathematics, University of Johannesburg, Auckland Park, 2006, South Africa}
 \affiliation{$^{b}$Quantum Research Group, School of Chemistry and Physics, University of KwaZulu-Natal, Durban, 4001, South Africa}
  \affiliation{$^{c}$National Institute for Theoretical Physics (NITheP), KwaZulu-Natal, 4001, South Africa}
 \affiliation{$^{d}$School of Electrical Engineering, KAIST, Daejeon, 34141, Republic of Korea}

\begin{abstract}
Open quantum walks (OQWs) describe a quantum walker on an underlying graph whose dynamics is purely driven by dissipation and decoherence. Mathematically, they are formulated as completely positive trace preserving (CPTP) maps on the space of density matrices for the walker on the graph. Any microscopically derived OQW must include the possibility of remaining on the same site on the graph when the map is applied. We extend the CPTP map to describe a lazy OQW. We derive a central limit theorem for lazy OQWs on a $d$-dimensional lattice, where the distribution converges to a Gaussian. We show that the properties of this Gaussian computed using conventional methods agree with the general formulas derived from our central limit theorem.
\end{abstract}

\maketitle

\section{Introduction}
Random walks have been applied to a vast number of areas in science including physics, computer science, financial economics and biology \cite{barber1970random,10.2307/1993657,inbook,papadimitriou94,cootner1964random,berg1993random}. Elevating the random walk onto the quantum level was first performed in the context of closed systems undergoing unitary evolution. Models for unitary quantum walks in discrete and continuous time have been proposed in \cite{PhysRevA.48.1687} and \cite{PhysRevA.58.915}, respectively. Comprehensive overviews for some of these early quantum walk models can be found in \cite{doi:10.1080/00107151031000110776} and \cite{VenegasAndraca:2012fh}. These models are comprised of a walker on an underlying graph. The walker possesses internal degrees of freedom (for example spin or polarization) which play a non-trivial role in determining the probability distribution on the graph. The unitary operator driving the evolution performs a transformation of the walker's internal degrees of freedom and then, depending on this resulting state, shifts the walker from one position on the graph to another. This unitary operator is applied at each time step and a coherent superposition between all the possible positions emerges.

It is well known that quantum walks display very different behaviour compared to their classical counterparts. In particular, it is interesting to study and compare the asymptotic, or long time, behaviour of the walks. The unitary quantum walks propagate outwards from the initial position quadratically faster than the classical random walk. A central limit theorem was derived in \cite{konno2005} in which the limit distribution was found not to be Gaussian, as it is for the classical case. Instead the distribution density was a function of the form
\begin{eqnarray}
	f\left(x\right) = \frac{\sqrt{1 - \left|a\right|^{2} }\left( 1 - \lambda x \right)}{\pi\left( 1-x^{2} \right)\sqrt{\left|a\right|^{2} - x^{2}}},
\end{eqnarray}
where $\lambda$ and $a$ are constants. Quantum walks have proved to be important when designing algorithms to perform a variety of search related tasks \cite{2011AcPSl..61..603R}. A significant feature omitted in these models is an additional non-zero probability for the walker to remain on the same site on the graph. In \cite{Childs2010} a lazy unitary quantum walk was constructed to establish a relationship between discrete time and continuous time quantum walks in certain limits. Grover's search algorithm \cite{10.1145/237814.237866} was formulated as a lazy unitary quantum walk in \cite{2015JPhA48Q5304W}. For the discrete quantum walk case, adding additional self-loops at the graph vertices, i.e., increasing the probability of the walker staying put, was found to affect the performance of the search algorithm, either improving the success probability (for a single self-loop) or hindering it (for more than one self-loop). The continuous quantum walk case was also studied with the number of self-loops having no affect. Aspects of probability distributions for lazy quantum walks were studied in \cite{PhysRevE.72.056112,PhysRevA.90.012342}. One notable feature to have been discovered is that of localization. Depending on the unitary evolution operator possessing eigenvalues isolated from its continuous spectrum, the walker exhibits a non-vanishing probability to remain at any position on its underlying graph. The lazy quantum walk model constructed in \cite{1674-1056-24-5-050305} used a $3\times 3$ Discrete Fourier Transform as its coin operator and compared its probability distribution to that of the non-lazy quantum walk with the usual $2\times 2$ Hadamard coin. The two distributions are different but have very similar probability distribution concentrated intervals. Furthermore, the $n$th moment of the two respective probability distributions were both shown to be order $n$ in time. It was also learned that lazy quantum walks have higher occupancy rate than other walks, classical or quantum. Improvements and further discoveries for the above models have been found in \cite{PhysRevA.100.042303, Giri_2019, Wong_2018}. Quantum walks have by now extensively been demonstrated experimentally. See for example the recent work of \cite{2019npjQI...5...40S} and references therein.

Unitary evolution is indicative of a closed quantum system. In this work, we will be concerned with a discrete time open quantum system random walk model. One in which the walk is driven by a dissipative environment. Open system quantum walk models were first introduced in \cite{OQWs1, ATTAL20121545, OQWs3}. These open quantum walks (OQWs) describe a system comprising of the walker possessing internal degrees of freedom and the underlying graph. The evolution of the walker is driven by a dissipative environment, where the interaction with this environment takes place between any two connected nodes. These non-unitary dynamics are described mathematically by completely positive trace preserving (CPTP) maps. These maps transform the internal degrees of freedom while shifting the walker from one position on the graph to another, thus again building up a statistical mixture of terms for each possible position contributing to the system's density matrix. The probability distribution of the walker's position for large times converges to a Gaussian, reminiscent of the classical random walk behaviour. 

With an appropriate choice of map, the OQW reproduces the classical Markov chain. A `physical realisation' procedure establishes a relation between the OQW and the unitary quantum walk \cite{OQWs1}. An OQW formulation of dissipative quantum computing (DQC) was presented in \cite{Sinayskiy2012}, in which the OQW based algorithms converged faster to the desired steady state, and had a higher probability of detection, than the canonical DQC models. Furthermore, the OQW allows for a quantum trajectory \cite{0305-4470-37-49-008} description which, in turn, allows for a quantitative analysis of the long-time, or asymptotic, behaviour of the OQW. Using quantum trajectories the work of \cite{2012arXiv1206.1472A} formulated a central limit theorem (CLT) for the discrete time homogeneous OQW where the underlying graph is a lattice $\mathbb{Z}^{d}$. They further managed to derive an explicit formula for the variance of the corresponding Gaussian. Using the CLT  \cite{2012arXiv1206.1472A}, the work of \cite{2013JSPKonno} introduced a Fourier space dual process for the OQWs and from this, they were able to find formal expressions for the probability distribution and, for a range of OQWs, the mean and variance for the corresponding distributions. Continuous time OQWs were first formulated in \cite{Pellegrini2014}, and the CLT was proved in \cite{Bringuier2017}. The authors of \cite{Sadowski2016} managed to generalize the CLT to some particular non-homogenous cases of the OQW on the lattice. Next, \cite{1402-4896-2012-T151-014077} studied the asymptotic probability distributions for OQWs on $\mathbb{Z}$ where the operators in the CPTP map are simultaneously diagonalizable. The asymptotic distributions were found to consist of, at most, two soliton-like solutions along with a certain number of Gaussians. Furthermore, they uncovered connections between the spectrum of the operators and properties of the asymptotic distributions. As will be elucidated below, the OQW quantum trajectories may be seen as classical Markov chains. Indeed many notions present in classical Markov chain theory, such as irreducibility, period and communicating classes, have been successfully introduced to OQWs through the quantum trajectory route \cite{Carbone2016, Carbone2015}, and the notion of hitting time for the OQW was defined in \cite{Lardizabal2016}. Applying the generic results of \cite{Carbone2016} to homogeneous OQWs on $\mathbb{Z}^{d}$, \cite{Carbone2015} proved the CLT as well as formulated the large deviation principle for quantum trajectories for OQWs.

Lastly, in the scaling limit, OQWs gave rise to a new class of Brownian motion, namely, Open Quantum Brownian Motion \cite{PhysRevA.88.062340, 1742-5468-2014-9-P09001}. These models do not exhibit Gaussian behaviour and no CLT is yet known. The detailed account of current status of the field of OQWs can be found in  \cite{SinayskiyEPJ2019}.

As with the unitary quantum walk case previously, a significant feature omitted thus far in the OQW model described above is the possibility of the walker to remain on the same site after the CPTP map is applied. After the initial model was proposed in \cite{OQWs1, ATTAL20121545, OQWs3} and the CLT derived in \cite{2012arXiv1206.1472A}, the
OQW was derived from a microscopic model \cite{PhysRevA.92.032105} in which the system and the environment, concretely  chosen  to  be  a  bath  of  harmonic  oscillators, together  constituted  a  closed system.  After a quantum master equation was derived for the system's reduced density matrix, the discrete time OQW was then obtained through a discretisation procedure. Crucially, the work of \cite{PhysRevA.92.032105} explicitly demonstrates that all microscopically derived OQWs must necessarily have a self-jumping term. Thus, any OQW model that is derived from a microscopic approach must incorporate the possibility of the walker remaining on the same site. We will call such a model a lazy open quantum walk. While the results derived in the above-mentioned references for the non-lazy OQWs, for example the central limit theorem, are significant the issue of deriving analogous results for a lazy OQW model remained an important and open physical problem. Thus, motivated by the results and insights of \cite{PhysRevA.92.032105} we extend the CPTP map to include an additional operator to encode for the possibility of a lazy open quantum walker. This then raises an interesting question about the long-time behaviour of the new lazy OQW model. In this work, we extend the central limit theorem of \cite{2012arXiv1206.1472A} to the lazy discrete OQW. This will fill an important gap in the OQW literature since the work of \cite{PhysRevA.92.032105}.

This paper is structured as follows. In section 2, we describe the discrete time homogenous lazy open quantum walk on the $d$-dimensional lattice, $\mathbb{R}^{d}$. We also introduce the Markov chain, through the quantum trajectory description, that will allow for the formulation of the new extended CLT. In section 3, we discuss the CLT for our lazy OQW, revise some important aspects of the microscopic derivation, and then connect the homogeneous OQW on the lattice to the quantities in the microscopic derivation. Lastly in this section, we study some examples in which we conduct non-trivial checks of the variance formula obtained from the CLT. Lastly, in section 4 we conclude our findings and identify some outstanding problems still present in the literature.

\section{Lazy open quantum walk formulation}
\subsection{The basic formulation}\label{sec:1}

We first introduce the OQW on a general graph. The graph consists of a set of nodes $\mathcal{V}$ and we define the set of all oriented edges on the graph $\{ \left( i,j \right) | i,j \in \mathcal{V} \}$. These oriented edges denote the possible transitions between the nodes in $\mathcal{V}$. Let the total number of nodes be $P$, where $P$ can either be finite or countably infinite. The Hilbert space consisting of states describing the position of the walker on the graph is $\mathcal{K} = \mathbb{C}^{P}$ for finite $P$, and $\mathcal{K} = l^{2}\left( \mathbb{C} \right)$ for $P$ being infinite. Here, $l^{2}\left( \mathbb{C} \right)$ is the space of square integrable functions. We will denote the orthonormal basis for $\mathcal{K}$ by $\ket{i}$, where $i \in \mathcal{V}$. The walker on this graph posseses internal degrees of freedom described by an $n$-dimensional Hilbert space $\mathcal{H}$. These internal degrees of freedom could represent spin, or polarisation or energy levels. The state of the walker's internal degree of freedom is given by the operator $\tau \in \mathcal{B}\left(\mathcal{H}\right)$, where $\mathcal{B}\left(\mathcal{H}\right)$ is the space of bounded operators acting on $\mathcal{H}$. To specify the state of the quantum walker, we need to specify its internal state and its position on the graph. The total state of the system is thus given by a density matrix $\rho$ on the tensor product space of $\mathcal{H}\otimes \mathcal{K}$. Thus, we have $\rho \in \mathcal{B}\left(\mathcal{H}\otimes \mathcal{K}\right)$.

We want the walk on the graph to be driven by dissipation. Between each two connected nodes on the graph, we envisage an external environment, for example a heat bath, interacting with the system. We define generalized quantum coin operators $B^{i}_{j}$ (in comparison to the unitary quantum walk coin) which describes dissipative interaction between the nodes $j$ and $i$. These operators satisfy the normalization condition 
\begin{equation}
 \sum\limits_{i\in \mathcal{V}}B^{i}_{j}\,\!^{\dagger} B^{i}_{j} = I.
\label{eq:Probconscond}
\end{equation}
This, in turn, defines a completely positive trace preserving map in Kraus representation \cite{1983LNP...190.....K}
\begin{equation}
	\mathcal{M}_{j}\left(\tau\right) = \sum\limits_{i\in \mathcal{V}}  B^{i}_{j}\tau B^{i}_{j}\,\!^{\dagger}.
\label{eq:KrausRep0}
 \end{equation}
For each node $j$, $\mathcal{M}_{j}$ serves to transform the internal degree of freedom and equation (\ref{eq:Probconscond}) ensures probability conservation. Figure \ref{fig:Fig1} shows an illustration of an OQW on a graph. Three nodes on the graph are labelled $i,j,k$ and the operators $B^{i}_{j}$, for example, describe the transformation of the walker's internal degree of freedom as a transition from site $j$ to site $i$ is made. 
\begin{widetext}
\begin{center}
\begin{figure}
\centering
   \includegraphics[width=.75\textwidth]{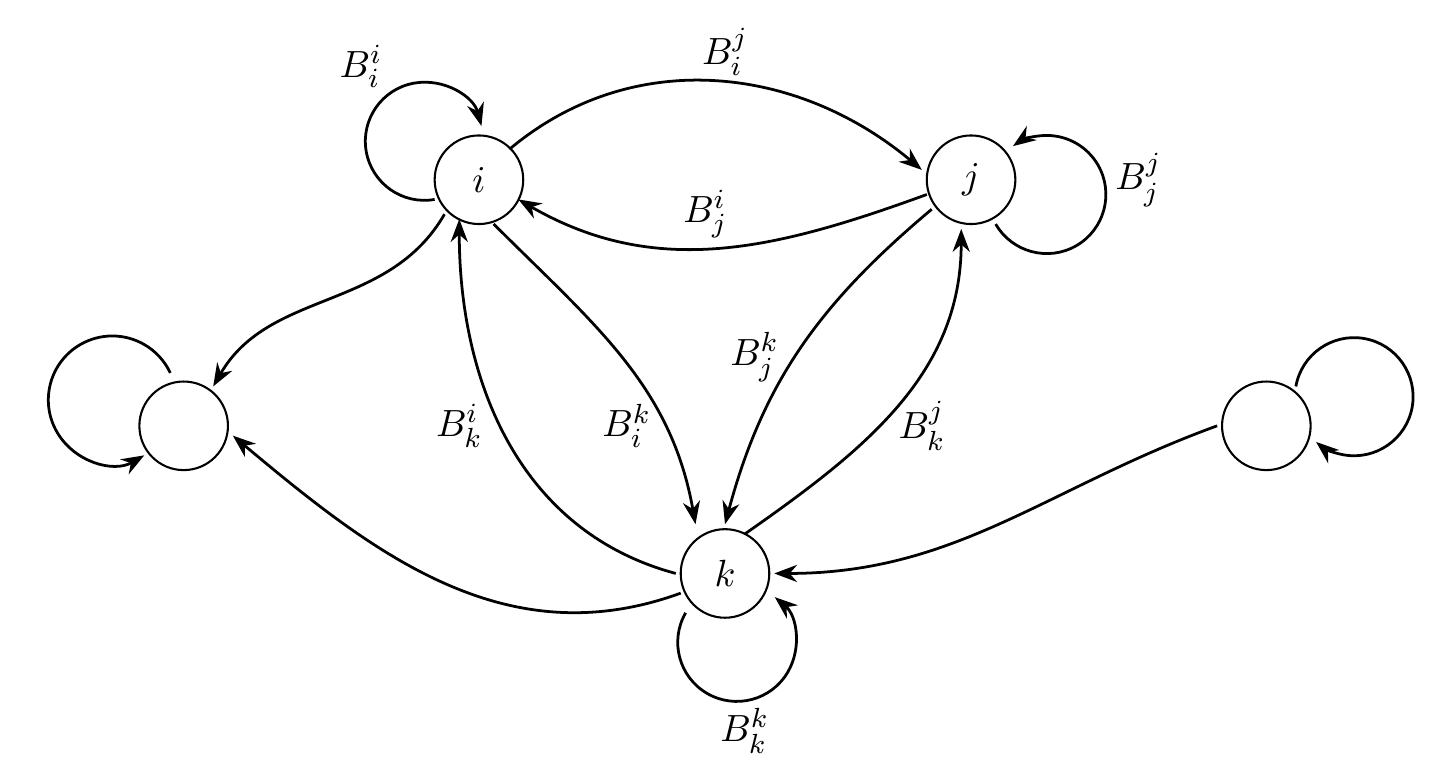}
  \caption{The above figure is an illustration of the lazy OQW. Three sample nodes, labeled $i,j$ and $k$, are shown for the underlying graph. The transition from node $j$ to node $i$, for example, represented by the directed arrow between those two nodes is described by the $B^{i}_{j}$ operator. $B^{i}_{j}$ transforms of the walker's internal degree of freedom as a transition from site $j$ to site $i$ is made. Since this is a lazy OQW, the walker also has the possibility of remaining on the same node. The operator $B^{j}_{j}$ encodes for this possibility, transforming the internal degree of freedom when the walker remains on site $j$.}
  \label{fig:Fig1}
\end{figure}
\end{center}
\end{widetext}
So far, we have only described the dynamics on the space $\mathcal{H}$. To formulate the jumping process, and thus describe the dynamics on the full tensor product space, we extend the map $\mathcal{M}$ on $\mathcal{B}\left(\mathcal{H}\right)$ to a map on $\mathcal{B}\left(\mathcal{H}\otimes \mathcal{K}\right)$. We define
\begin{equation}
	M^{i}_{j} = B^{i}_{j} \otimes \ket{i}\!\!\bra{j}, \hspace{20pt}\sum\limits_{i\in \mathcal{V}} \sum\limits_{j\in \mathcal{V}} M^{i}_{j}\,^{\dagger}M^{i}_{j} = I,
 \label{eq:MaponFullspace}
 \end{equation}
 where the identity operator here is defined on $\mathcal{H}\otimes \mathcal{K}$. We can now define a CPTP map on the density matrix $\rho \in \mathcal{B}\left(\mathcal{H}\otimes \mathcal{K}\right)$ 
\begin{equation}
	\mathcal{M}\left(\rho\right) = \sum\limits_{i\in \mathcal{V}} \sum\limits_{j\in \mathcal{V}} M^{i}_{j}\rho M^{i}_{j}\,\!^{\dagger}.
\label{eq:KrausRep12}
\end{equation}
The CPTP map constitutes a discrete time open quantum walk on a graph. Starting from an arbitrary initial state at time $t=0 $ say, $ \rho^{\left(0\right)} = \sum_{i,j}  \tau^{\left(0\right)}_{i,j}\otimes \ket{i}\!\!\bra{j}$, one can show that the form of the state becomes diagonal in the position space $\mathcal{K}$ after a single application of $\mathcal{M}$:
\begin{eqnarray}
	\mathcal{M}\left(\rho^{\left(0\right)}\right) &=& \sum\limits_{i\in \mathcal{V}} \sum\limits_{j\in \mathcal{V}} M^{i}_{j}\left( \sum\limits_{k\in \mathcal{V}} \sum\limits_{l\in \mathcal{V}}  \tau^{\left(0\right)}_{k,l} \otimes \ket{k}\!\!\bra{l} \right) M^{i}_{j}\,^{\dagger}\\
	&=&  \sum\limits_{i\in \mathcal{V}} \left(  \sum\limits_{j\in \mathcal{V}} B^{i}_{j}\tau^{\left(0\right)}_{j,j} B^{i}_{j}\,^{\dagger}  \right) \otimes \ket{i}\!\!\bra{i}.
\label{eq:diagonal}
\end{eqnarray}
The density matrix at time $t = 1$ then has the form
\begin{equation}
	\rho^{\left(1\right)} = \sum\limits_{i\in \mathcal{V}} \tau^{\left(1\right)}_{i} \otimes  \ket{i}\!\!\bra{i}, \hspace{15pt} \tau^{\left(1\right)}_{i} =  \sum\limits_{j\in \mathcal{V}} B^{i}_{j}\tau^{\left(0\right)}_{j,j} B^{i}_{j}\,^{\dagger}.
\label{eq:diagonal2}
\end{equation}
This indicates that there is no mixing taking place between the positions on the graph in our OQW model. For this reason we restrict our attention to density matrices of the form $\rho = \sum\limits_{i} \tau_{i} \otimes \ket{i}\!\!\bra{i}$. The density matrix at time $t = n$ may be obtained through iteration:
\begin{equation}
	\rho^{\left(n\right)} = \sum\limits_{i\in \mathcal{V}} \tau^{\left(n\right)}_{i} \otimes  \ket{i}\!\!\bra{i}, \hspace{15pt} \tau^{\left(n\right)}_{i} =  \sum\limits_{j\in \mathcal{V}} B^{i}_{j}\tau^{\left(n-1\right)}_{j} B^{i}_{j}\,^{\dagger}.
\label{eq:diagonaln}
\end{equation}
The probability that a position measurement, at time $t=n$, will yield a result of $X_{n} = i$ is $p\left(X_{n} = i\right) = \mathrm{Tr}( \tau^{\left(n\right)}_{i})$, with the sum over all the positions $i$ in $p\left(X_{n} = i\right)$ equal to 1. For a more comprehensive introduction to OQWs on graphs, see \cite{OQWs1,ATTAL20121545, OQWs3}. 
\begin{center}
\begin{figure}
\centering
   \includegraphics[width=.45\textwidth]{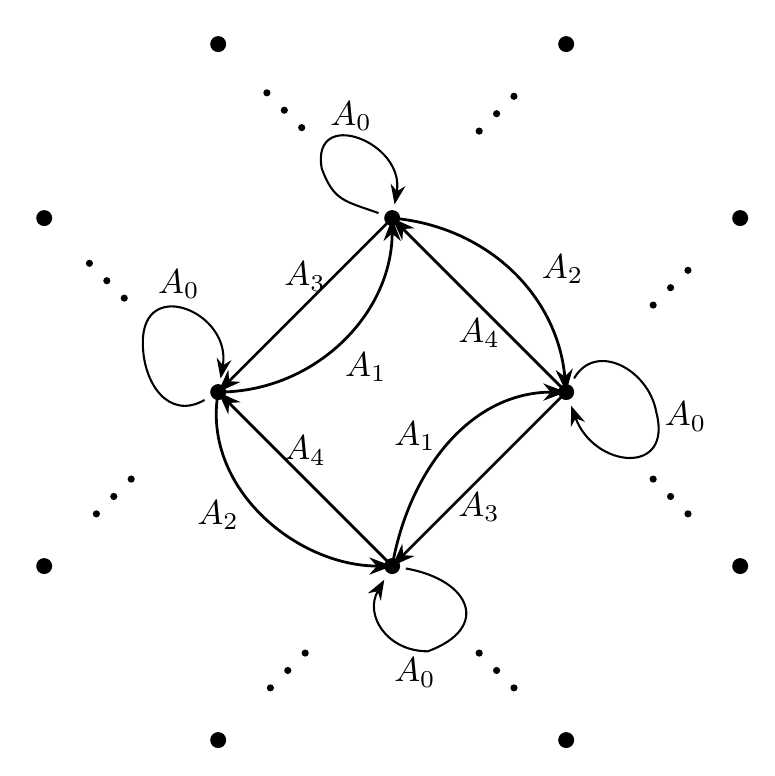}
  \caption{The above figure is an illustration of the homogeneous lazy OQW model on a two-dimensional lattice. A representative sample of four lattice points are shown here. In this case, $d=2$. The canonical basis elements are $\{ e_{1},e_{2},e_{3},e_{4} \}$ where $e_{1}$ and $e_{2}$ indicate the two positive directions. The elements $e_{3} = -e_{1}$ and $e_{4} = -e_{2}$ indicate the corresponding negative directions respectively. Transitions along the positive directions are described by operators $A_{1}$ and $A_{2}$ respectively. The operators encoding for transitions in the corresponding negative directions are $A_{3}$ and $A_{4}$. At every node in the lattice there is the presence of a self-jumping arrow represented by an extra operator $A_{0}$.}
  \label{fig:Fig2}
\end{figure}
\end{center}
For the rest of this work, we will consider a homogeneous (translation invariant) discrete time open quantum walk whose underlying graph is a $d$-dimensional lattice. We employ the use of the canonical basis $\{e_{1}, e_{2}, ..., e_{d} \}$ on $\mathbb{Z}^{d}$, with $e_{d+j} = -e_{j}$ for all $j = 1, ..., d$. Thus, for each site on the lattice there are $2d$ adjacent sites for the walker to jump to - one for each direction corresponding to the $e_{i}$'s. We also will include $e_{0} = 0$ in our formulation to encode the idea that the walker can remain on the same site. Then in total there are $2d+1$ possible jumps. For a homogeneous walk, all of the $B$ operators along the positive $i$th direction in $\mathbb{Z}^{d}$ are identical and are denoted by $A_{i}$, while all $B$'s along the negative $i$th direction will be denoted by $A_{i+d}$. The precise relation between the $B$ operators and the $A$ operators is, for the $i$th direction 
\begin{equation}
B^{k+1}_{k} = A_{i}, \hspace{20pt} B^{k-1}_{k} = A_{i+d}, \hspace{20pt}\forall \;k \in \mathbb{Z}.
\label{eq:relAsandBs}
\end{equation}
Here, $k$ labels the nodes in the $e_{i}$ direction. The position space of the walker is the Hilbert space $\mathcal{K} = \mathbb{C}^{\mathbb{Z}^{d}}$, the basis for which is denoted by $\left(\ket{i}\right)_{i\in \mathbb{Z}^{d}}$. In previous formulations of open quantum walks on the lattice, a family of bounded operators $\{A_{1},..., A_{2d}\} \in \mathcal{B}\left(\mathcal{H}\right)$ performed transformations on the state $\tau$ as the walker necessarily jumped to an adjacent site. 
We extend the family of bounded operators to include an extra operator $A_{0}$ representing the effect of remaining on the same site. An illustration of this model is shown in Figure \ref{fig:Fig2}. Thus we have $\{A_{0}, A_{1},..., A_{2d} \}$ acting on $\mathcal{H}$ and satisfying 
\begin{equation}
 \sum\limits_{j=0}^{2d} A^{\dagger}_{j} A_{j} = I.
 \label{eq:normalisation}
 \end{equation}
The completely positive map on $\mathcal{B}\left(\mathcal{H}\right)$ in the Kraus representation is now
\begin{equation}
	\mathcal{L}\left(\tau\right) = \sum\limits^{2d}_{j=0} A_{j}\tau A^{\dagger}_{j}.
\label{eq:KrausRep}
 \end{equation}
We extend the map from $\mathcal{B}\left(\mathcal{H}\right)$ to $\mathcal{B}\left(\mathcal{H}\otimes \mathcal{K}\right)$ with
\begin{equation}
	M^{j}_{i} = A_{j} \otimes \ket{i+e_{j}}\!\!\bra{i}.
\label{eqn:CPTPmap}
 \end{equation}
The operator acting on $\mathcal{K}$ in (\ref{eqn:CPTPmap}) describes the transition from lattice site $i$ either to the adjacent site in the $j$th direction for $j = 1,2,..., 2d$, or to the same site $i$ again for $j=0$. We still have 
\begin{equation}
	\sum\limits_{i\in \mathbb{Z}^{d}}\sum\limits^{2d}_{j=0} \left(M^{j}_{i} \right)^{\dagger} \left( M^{j}_{i} \right) = I,
 \end{equation}
where $I$ here is on $\mathcal{H}\otimes \mathcal{K}$. The CPTP map, defining the discrete time homogeneous open quantum walk is
\begin{equation}
	\mathcal{M}\left( \rho \right) = \sum\limits_{i\in \mathbb{Z}^{d}}\sum\limits^{2d}_{j=0} M^{j}_{i}\rho \left(M^{j}_{i}\right)^{\dagger}.
 \end{equation}
We will still be interested in density matrices of the form $\rho = \sum_{i\in \mathbb{Z}^{d}}\tau_{i} \otimes \ket{i}\!\!\bra{i}$. For $\rho$ to be normalized, we must have $\sum_{i\in \mathbb{Z}^{d}} Tr \left( \tau_{i} \right) =1$. If, at time $n$, the state of the system is
\begin{equation}
	\rho^{\left(n\right)} = \sum\limits_{i\in \mathbb{Z}^{d}}\tau^{\left(n\right)}_{i} \otimes \ket{i}\!\!\bra{i},
\end{equation}
then, after applying $\mathcal{M}$, the state at time $n+1$ is
\begin{subequations}
\begin{eqnarray}
	\rho^{\left(n+1\right)} &=& \mathcal{M}\left(\rho^{\left(n\right)}\right) =  \sum\limits_{i\in \mathbb{Z}^{d}}\tau^{\left(n+1\right)}_{i} \otimes \ket{i}\!\!\bra{i},\\ \tau^{\left(n+1\right)}_{i} &=& \sum\limits^{2d}_{j = 0} A_{j}\tau^{\left(n\right)}_{i-e_{j}}A^{\dagger}_{j}.
 \end{eqnarray}
 \end{subequations}
A very important ingredient in our formulation of the central limit theorem is the steady state $\rho_{\infty} \in \mathcal{B}(\mathcal{H})$, defined to be invariant under the CPTP map in (\ref{eq:KrausRep}) $\rho_{\infty} = \mathcal{L}\left(\rho_{\infty}\right)$. 
In our formulation of the lazy OQW $\rho_{\infty}$, due the inclusion of the $A_{0}$ term in $\mathcal{L}$, will indeed be different from the previous OQWs where staying on the same site was not possible. We now present a particular example of a lazy OQW. Consider a lazy walk on the line $(d=1)$ with a two-dimensional coin space. There will be three matrices $A_{1}, A_{0}$ and $A_{2}$ for moving forward, staying on the same site, and moving backwards, respectively. Figure \ref{fig:Fig3} shows an illustration of this particular OQW in two dimensions. In this example, we take
\newpage
\begin{eqnarray}
	&&A_{1} = \frac{1}{\sqrt{6}}\left(\begin{array}{cc}1 & 1 \\1 & e^{i\pi/3}\end{array}\right), \hspace{10pt} A_{0} =  \frac{1}{\sqrt{6}}\left(\begin{array}{cc}1 & e^{2i\pi/3} \\1 & -1\end{array}\right),\nonumber \\
 &&A_{2} = \frac{1}{\sqrt{6}}\left(\begin{array}{cc}1 & e^{-2i\pi/3} \\1 & e^{-i\pi/3}\end{array}\right).
\label{eq:ExampleOQW}
\end{eqnarray}
Indeed, these operators satisfy $A_{1}^{\dagger}A_{1} + A_{0}^{\dagger}A_{0} + A_{2}^{\dagger}A_{2} = I$. From Figure \ref{fig:Fig4}, with the initial state chosen to be $I/2\otimes \ket{0}\!\!\bra{0}$, one can see that the probability distribution of the lazy OQW approaches a Gaussian distribution centred at the origin.
\begin{widetext}
\begin{center}
\begin{figure}
\centering
   \includegraphics[width=.90\textwidth]{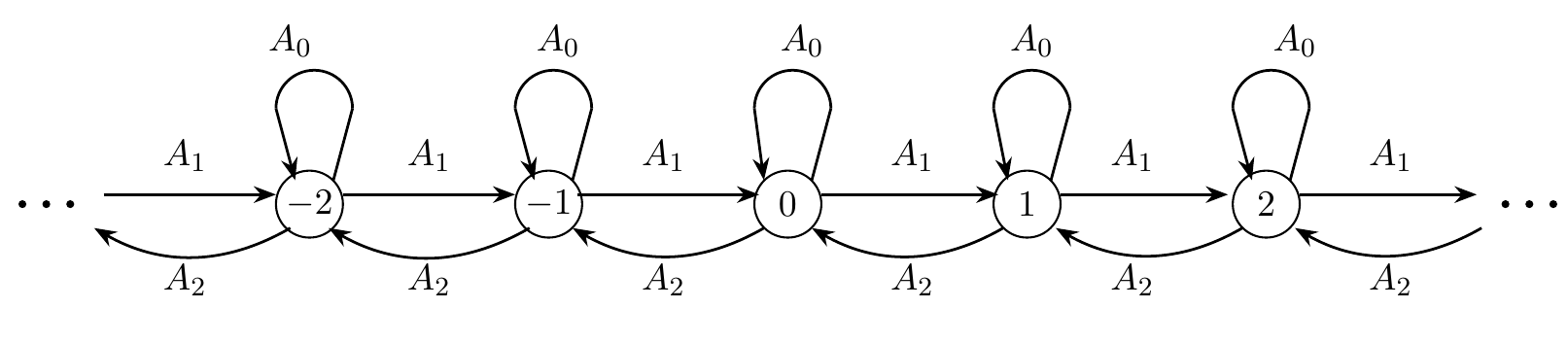}
  \caption{This figure depicts the discrete homogeneous lazy OQW on the line. The operators $A_{1}, A_{2}$ shift the walker forwards and backwards respectively, while $A_{0}$ corresponds to remaining on the same site.}
  \label{fig:Fig3}
\end{figure}
\end{center}
\end{widetext}
\begin{widetext}
\begin{figure*}
\centering
    \includegraphics[width=.85\textwidth]{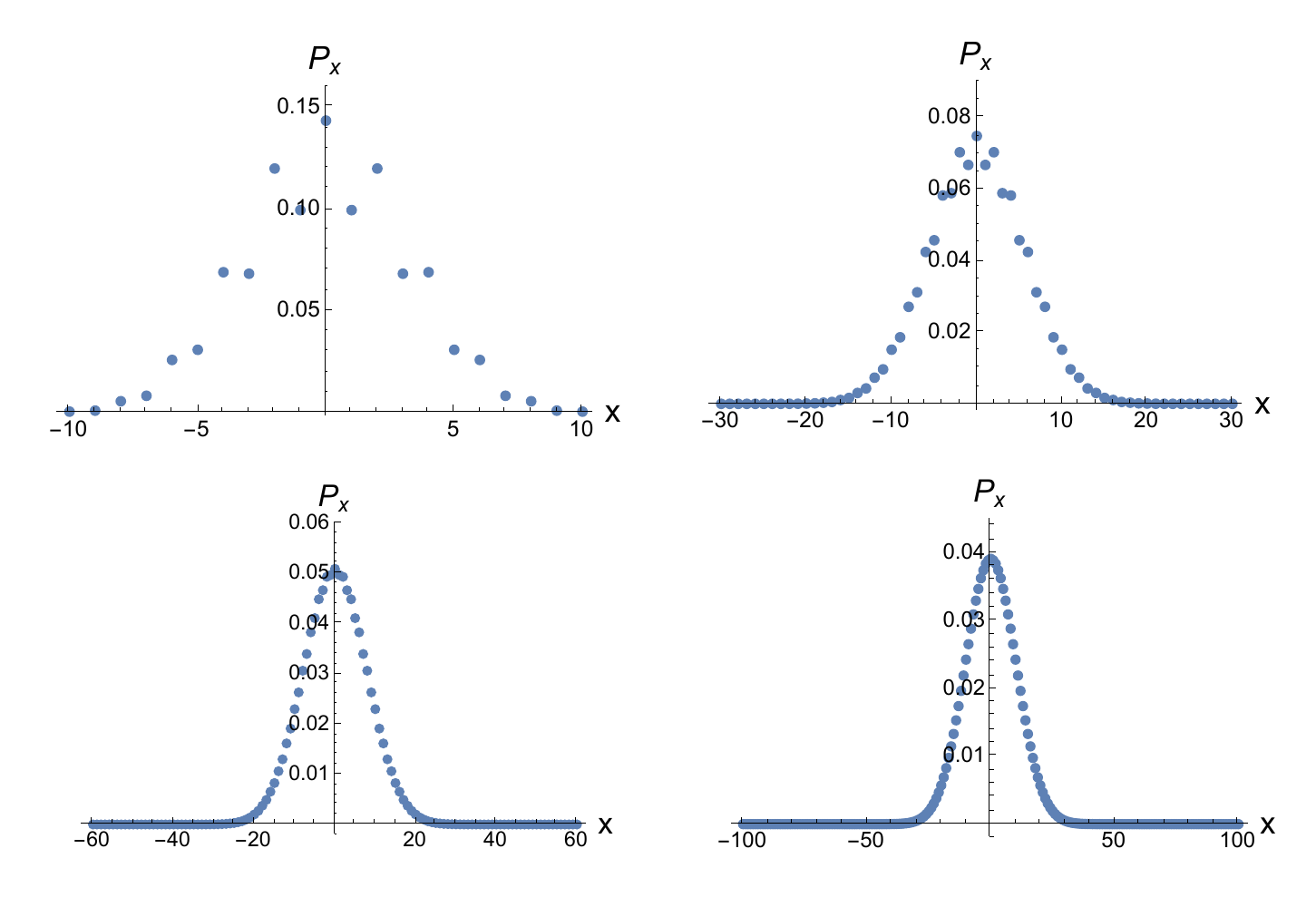}  
 \caption{In these four figures, we ran the OQW for the operators defined in equation (\ref{eq:ExampleOQW}) for $10, 30, 60$ and then $100$ successive steps. The horizontal axis label x labels position on the $x$-axis, while $P_{x}$ on the vertical axis denotes the probability. Top left: for $n=10$. Top right: for $n=30$. Bottom left: for $n=60$. Bottom right: for $n=100$. The non-lazy version of the OQW (where the walker had no option to remain on the same site) also exhibited Gaussian behaviour. One way to understand this was due to the decoherence of the state after a single application of the OQW map. As can be seen from equation (\ref{eq:diagonal2}), any mixing between different positions on the graph vanishes.}
  \label{fig:Fig4}
\end{figure*}
\end{widetext}

\subsection{Quantum Trajectories}\label{sec:2}
In this section we summarize a generic unraveling of OQWs in the quantum trajectory frame introduced in \cite{2012arXiv1206.1472A}. Quantum trajectories are a convenient way to simulate the OQW. Furthermore, the CLT formulations of OQWs thus far have all been within the quantum trajectory framework. Beginning from the general formulation of the OQW, the idea for the quantum trajectory is the following. Consider some state at time $t = n$, $\rho^{(n)} = \tau_{n} \otimes \ket{i_{n}}\!\!\bra{i_{n}}$, where the position of the walker is $X_{n} = i_{n}$. We apply the map $\mathcal{M}$ and then perform a measurement on the position space $\mathcal{K}$. The state then, at the time $n+1$, jumps to
\begin{subequations}
\begin{eqnarray}
	\rho^{(n+1)} &=& \frac{1}{\mathbb{P}\left(i_{n+1}| i_{n}\right)}B^{i_{n+1}}_{i_{n}}\tau_{n}\left(B^{i_{n+1}}_{i_{n}}\right)^{\dagger} \otimes \ket{i_{n+1}}\!\!\bra{i_{n+1}},\nonumber \\
	\label{eq:QT1}\\
	\mathbb{P}\left(i_{n+1}|i_{n}\right) &=& \mathbb{P}\left(X_{n+1} = i_{n+1}|X_{n} = i_{n}\right)\nonumber \\
	&=& \mathrm{Tr}\left( B^{i_{n+1}}_{i_{n}}\tau_{n}\left(B^{i_{n+1}}_{i_{n}}\right)^{\dagger}  \right)
	\label{eq:QT2}
\end{eqnarray}
\end{subequations}
is the conditional probability of the position $X$ at time $n+1$ being equal to $i_{n+1}$ given that $X$ at time $n$ was $i_{n}$. Repetition of this process leads to a classical Markov chain valued in the set of states of the form $\tau \otimes \ket{i}\!\!\bra{i}$. One may denote this Markov chain as $\left( \tau_{n} , X_{n} \right)_{n \geq 0}$. Thus, working in the quantum trajectory framework allows us to consider a classical Markov chain and thus may be susceptible to standard theorems in probability theory literature. In particular the central limit theorem is of particular relevance in this work. Averaging over this quantum trajectory procedure simulates an OQW Master equation driven by $\mathcal{M}$ as can be seen from
\begin{eqnarray}
	\mathcal{E}\left( \rho^{(n+1)} \right) &=& \sum\limits_{i_{n+1}}\mathbb{P}\left(i_{n+1}|i_{n}\right)\rho^{(n+1)}\nonumber
	\\
	&=& \sum\limits_{i_{n+1}} B^{i_{n+1}}_{i_{n}}\tau_{n}\left(B^{i_{n+1}}_{i_{n}}\right)^{\dagger} \otimes \ket{i_{n+1}}\!\!\bra{i_{n+1}}\nonumber\\
	&=& \mathcal{M}\left(\rho^{(n)} \right).
	\label{eq:QuantTrajsimOQW}
\end{eqnarray}
Above we used equation (\ref{eq:QT1}) for $\rho^{(n+1)}$. Extending the map to include $A_{0}$ also admits a quantum trajectory description with a Markov chain $\left( \tau_{n} , X_{n}\right)$. For our homogeneous OQW on the lattice, the quantum trajectory description is the following. Let the state of the system at time $t = n$ be $\left( \tau_{n} , X_{n} = i\right) $. Apply the open quantum walk map $\mathcal{M}$ to the state performing a measurement of the position directly after. The state then at time $t = n+1$ jumps to
\begin{eqnarray}
	\left(\frac{1}{\mathbb{P}\left( j,n\right)}A_{j}\tau_{n}A^{\dagger}_{j} ,X_{n+1} = i+e_{j}\right),
\end{eqnarray}
with the probability $\mathbb{P}\left(j,n\right) = \Tr\left(A_{j}\tau_{n}A^{\dagger}_{j}\right)$. Note that even if the walker does remain on the same site, the probability for which would be $\mathbb{P}\left(0,n\right) = \Tr\left( A_{0}\tau_{n}A^{\dagger}_{0} \right)$, its state in $\mathcal{H}$ still undergoes a transformation by $\frac{1}{\mathbb{P}\left(0,n\right)}A_{0}\rho_{n}A^{\dagger}_{0}$. For quantum trajectories in discrete time, the sequence $\frac{1}{n}\sum\limits^{n}_{t=1}\tau_{t}$ converges almost surely to a random steady state $\rho_{\infty}$  \cite{0305-4470-37-49-008}: 
\begin{eqnarray}
	\frac{1}{n}\sum\limits^{n}_{j=1}\tau_{j} \overset{a.s }{ \longrightarrow}  \rho_{\infty}.
\label{eq:ergodicthass}
\end{eqnarray}
We further assume that the steady state $ \rho_{\infty}$ is unique.

The central limit theorem formulated in the following section will be formulated in terms of the random variables $\left( \tau_{n}, \Delta X_{n} \right)$, where $\Delta X_{n} = X_{n} - X_{n-1} \in \{ e_{0}, e_{1}, \cdots , e_{2d} \}$ and $n \neq 0$. This sequence $\left( \tau_{n}, \Delta X_{n} \right)_{n\geq 0}$ also forms a Markov chain. The transition operator from state $\left( \tau, e_{i} \right)$ to $\left( \tau', e_{j} \right)$ is given by
\begin{eqnarray}
	&&P\left[ \left( \tau, e_{i} \right), \left( \tau', e_{j} \right) \right] = \left\{\begin{array}{cc}\Tr\left( A_{j}\tau A^{\dagger}_{j} \right) & \mathrm{if} \; \tau' = \frac{A_{j}\tau A^{\dagger}_{j}}{\Tr\left( A_{j}\tau A^{\dagger}_{j} \right)} \\0 & \mathrm{otherwise.}\end{array}\right. \nonumber\\
\end{eqnarray}
The random variables that typically feature in central limit theorems are independent and identically distributed (iid). Once the OQW is in the steady state, the set of $\Delta X_{n}$ will constitute the iid random variables.

\section{The central limit theorem for the lazy walker}

\subsection{The central limit theorem}

A central limit theorem was proved for the open quantum walk in \cite{2012arXiv1206.1472A}, which we generalize to the lazy open quantum walk. Define the iid random variables $Y_{k} = \{e_{0}, e_{1}, \cdots e_{2d}\}$. For these random variables to be iid, we require the system to be in the steady state, $\rho_{\infty}$. We define the mean $m \in \mathbb{R}^{d}$ to be
\begin{eqnarray}
	m = \mathbb{E}\left(Y_{k}\right) = \sum\limits^{2d}_{i=0}\mathbb{P}\left( i \right)e_{i}, \hspace{7pt}  \mathbb{P}\left( i \right) = \Tr\left( A_{i}\rho_{\infty}A^{\dagger}_{i} \right).
\label{eq:mean}
\end{eqnarray}
The position of the walker on the lattice at time $n$ is
\begin{eqnarray}
	X_{n} = X_{0} + \sum\limits^{n}_{i=1}Y_{i},
\end{eqnarray}
where the initial position $X_{0}$ can be chosen for convenience to be zero. A central limit theorem can now be proven for the quantity
\begin{eqnarray}
	\frac{1}{\sqrt{n}}\left( X_{n} - \mathbb{E}\left(X_{n}\right) \right) &=& \frac{1}{\sqrt{n}}\left(X_{n}-nm\right) \nonumber \\
	 &=&\sqrt{n}\left( \frac{1}{n} \sum\limits^{n}_{i=1}Y_{i} - m \right).
\label{eqn:CLT1}
\end{eqnarray}
For any vector $l \in \mathbb{R}^{d}$, we may perform a Doob decomposition \cite{doob1953stochastic} of the quantity $\left(X_{n}-nm\right)\cdot l$. Essentially we are able to perform a decomposition of this quantity into a martingale $M_{n}$ and a predictable process $A_{n}$. The magnitude $|A_{n}|$, for $n$ taken over the positive integers, can be shown to be bounded independently of $n$ \cite{2012arXiv1206.1472A}. Thus, the process $A_{n}$ does not contribute to the law of large numbers and the central limit theorem at large times, i.e. for large $n$. The martingale is
\begin{eqnarray}
	M_{n} = \sum\limits^{n}_{j=2}\left[ f\left( \rho_{j} , \Delta X_{j} \right) - Pf\left( \rho_{j-1} , \Delta X_{j-1} \right)  \right],
\label{eq:martingale}
\end{eqnarray}
where the function $f$ is $f\left( \rho, x \right) = \Tr\left( \rho L_{l} \right) + x\cdot l$ and the operator $L_{l} = L \cdot l$ is a solution to the equation
\begin{subequations}
\begin{eqnarray}
	 \left( L_{l} - \mathcal{L}^{\dagger}\left( L_{l} \right) \right) &=& \sum\limits^{d}_{i=1}\widetilde{A}_{i}\left( e_{i} \cdot l \right) - \left( m \cdot l \right)I, \label{eq:TheLEq} \\ 
	\widetilde{A}_{i} &=& A^{\dagger}_{i}A_{i} - A^{\dagger}_{i+d}A_{i+d}.
	\label{eq:TheLEqb}
\end{eqnarray}
\end{subequations}
The quantity $M_{n}$ satisfies the defining condition for a martingale. Martingales feature prominently in probability theory \cite{hall1980martingale, williams1991probability}. One of the fundamental notions of probability theory is that of a $\sigma$-space spanned by events to which probabilities are assigned. In the current quantum trajectory setting, the events are $\left( \tau_{n} , X_{n} \right)$. We define the filtration $\left(\mathcal{F}_{n}\right)_{n\geqslant 2}$, where $\mathcal{F}_{n}$ is the $\sigma$-space spanned by events $\left( \tau_{j} , X_{j} \right)$ for $j \leqslant n $. The defining condition for martingale $M_{n}$ with respect to $\left(\mathcal{F}_{n}\right)_{n\geqslant 2}$ is
\begin{eqnarray}
	\mathbb{E}\left[ \Delta M_{n} | \mathcal{F}_{n-1} \right] = 0.
\label{eq:MartCond}
\end{eqnarray}
Note that the dual map $\mathcal{L}^{\dagger}$ in (\ref{eq:TheLEq}), defined as $\mathcal{L}^{\dagger}\left(\tau\right) = \sum^{2d}_{i=0}A^{\dagger}_{i}\tau A_{i},
$
is different for the lazy open quantum walk because of the inclusion of the $A_{0}$ operator. In this current work, we further extend analysis of equation (\ref{eq:TheLEq}) beyond that of \cite{2012arXiv1206.1472A} by noting that (\ref{eq:TheLEq}) forms a degenerate system of equations. One way of seeing this is by vectorising (\ref{eq:TheLEq}) with the help of the reshaping operation. The reshaping operation stacks the rows of a matrix on top of each other in a row vector. So for an $m\times n$ matrix $A$, for example, we have
\begin{eqnarray}
 &&   \mathrm{vec}\left(A\right)\\
      \label{eq:ResOp}
     && = \left( a_{11}, a_{12},\cdots, a_{1n}, a_{21}, \cdots, a_{2n}, \cdots , a_{m1}, a_{m2} , \cdots a_{mn}  \right)^{T}.\nonumber
\end{eqnarray}
The left-hand-side of (\ref{eq:TheLEq}) then becomes
\begin{eqnarray}
	\left( L_{l} - \mathcal{L}^{\dagger}\left( L_{l} \right) \right) \rightarrow \left(I - \sum\limits^{2d}_{i=1}A^{\dagger}_{i}\otimes A^{T}_{i}\right)\mathrm{vec}\left( L_{l} \right).
\label{eq:vectorisation}
\end{eqnarray} 
One can show, using (\ref{eq:normalisation}), that $\sum^{2d}_{i=1}A^{\dagger}_{i}\otimes A^{T}_{i}$ has an eigenvector of $\mathrm{vec}\left( I \right)$ with eigenvalue of 1. This means the determinant of the matrix in (\ref{eq:vectorisation}) vanishes, and the system of equations is degenerate. However, \cite{2012arXiv1206.1472A} proves that there exists a solution to (\ref{eq:TheLEq}) for any $l\in \mathbb{R}^{d}$, and that the difference between any two solutions is proportional to the identity. Furthermore, we note here that after taking the adjoint of (\ref{eq:TheLEq}), $L_{l} = L^{\dagger}_{l}$ is also a solution. Since $L_{l} + \alpha I$ also solves (\ref{eq:TheLEq}), we conclude that half of the $D\left(D-1\right)$ off-diagonal entries in $L_{l}$ are not independent, where $D$ is the dimension of the coin space, $\mathcal{H}$. Writing the vector $l$ in terms of the canonical basis $l = \sum^{d}_{i=1}l_{i} e_{i}$, we have $L_{l} = \sum^{d}_{i=1}L_{i}l_{i}$. It follows that there is an $L_{i}$ for each direction on the $d$-dimensional lattice satisfying
\begin{equation}
	L_{i} - \mathcal{L}^{\dagger}\left( L_{i} \right) = \widetilde{A}_{i} - m_{i}I.
\label{eq:eqforLi}
\end{equation}

The martingale in (\ref{eq:martingale}) satisfies the two necessary conditions for a central limit theorem for martingales to be applicable \cite{2012arXiv1206.1472A}, \cite{hall1980martingale}. The first condition is, for any $\epsilon > 0$,
\begin{eqnarray}
	\lim_{n\rightarrow +\infty} \frac{1}{n}\sum\limits^{n}_{k=1} \mathbb{E}\big[ 	(\Delta M_{k})^{2} \mathbb{I}_{|\Delta M_{k}|>\epsilon \sqrt{n}} \big] = 0
\label{eq:Martcond1}
\end{eqnarray}
where $\mathbb{I}_{|\Delta M_{k}|>\epsilon \sqrt{n}}$ is non-zero only when the condition $|\Delta M_{k}|>\epsilon \sqrt{n}$ is met. However, it is straightforward to show that $|\Delta M_{k}|$ is bounded above by a quantity independent of $k$ \cite{2012arXiv1206.1472A}. Thus, as $n$ continues to increase, there will come a point where the inequality changes to $|\Delta M_{k}| < \epsilon \sqrt{n}$. Thus, in the limit as $n \rightarrow \infty$ the $\mathbb{I}_{|\Delta M_{k}|>\epsilon \sqrt{n}}$ forces the entire expression to vanish. The second condition in the CLT is
\begin{eqnarray}
	\lim_{n\rightarrow +\infty} \frac{1}{n}\sum\limits^{n}_{k=1} \mathbb{E}\big[ 	(\Delta M_{k})^{2} | \mathcal{F}_{k-1} \big] = \sigma^{2}.
\label{eq:Martcond2}
\end{eqnarray}
Using equations (\ref{eq:martingale}), (\ref{eq:MartCond}) and the ergodic theorem for the unique $\rho_{\infty}$ (\ref{eq:ergodicthass}) one can show that the only surviving contribution to the left-hand side of (\ref{eq:Martcond2}) is indeed a finite quantity for each $l\in \mathbb{R}^{d}$ depending on the $m_{i}$, $L_{i}$, $\rho_{\infty}$ and the $A$ operators. With the two conditions for the central limit theorem satisfied, $M_{n}/\sqrt{n}$ converges in distribution to a Gaussian $\mathcal{N}\left(0,\sigma^{2}_{l}\right)$, where $\sigma^{2}_{l} = \sum^{d}_{i,j=1}l_{i}l_{j}C_{ij}$. It is remarkable that one may obtain an analytic expression for the covariance matrix of this distribution \cite{2012arXiv1206.1472A}
\begin{eqnarray}
	C_{ij} &=& \delta_{ij}\left( \Tr\left( A_{i}\rho_{\infty}A^{\dagger}_{i} \right) + \Tr\left( A_{i+d}\rho_{\infty}A^{\dagger}_{i+d} \right)	 \right) - m_{i}m_{j} \nonumber\\
	&&+\; \Big( \Tr\left( A_{i}\rho_{\infty}A^{\dagger}_{i}L_{j} \right) + \Tr\left( A_{j}\rho_{\infty}A^{\dagger}_{j} L_{i} \right) \nonumber \\
	&& - \Tr\left( A_{i+d}\rho_{\infty}A^{\dagger}_{i+d}L_{j} \right) - \Tr\left( A_{j+d}\rho_{\infty}A^{\dagger}_{j+d} L_{i}  \right) \Big) \nonumber \\
	&& - \left( m_{i}\Tr\left( \rho_{\infty}L_{j} \right) + m_{j}\Tr\left( \rho_{\infty}L_{i} \right) \right),
	\label{eqn:VarCij}
\end{eqnarray}
for each of the possible directions on the lattice $i,j = \{ 1,2, ,\cdots, d \}$. We have thus derived a central limit theorem for the homogeneous lazy OQW on a $d$-dimensional lattice. To calculate the covariance matrix using (\ref{eqn:VarCij}) we first need to calculate values for $m_{i}$, $L_{i}$ and $\rho_{\infty}$. The steady state is obtained by solving $\mathcal{L}(\rho_{\infty}) = \rho_{\infty}$. The lazy OQW map $\mathcal{L}$ is extended to include the $A_{0}$. Thus, solving this equation will yield a genuinely different steady state compared to the non-lazy OQW. The $m_{i}$ is calculated from the formula
\begin{equation}
	m_{i} = \Tr\left( \widetilde{A}_{i}\rho_{\infty}  \right)e_{i}
	\label{eq:formformj}
\end{equation}
where $\widetilde{A}_{i}$ is given by (\ref{eq:TheLEqb}). Finally the $L_{i}$ matrices are obtained by solving equation (\ref{eq:TheLEq}). This equation features the dual map $\mathcal{L}^{\dagger}$ which is also extended by the $A_{0}$ term, as well as the new $m_{i}$ values. Thus, the $L_{i}$ matrices will also be different from the non-lazy OQW. In what follows, we subject the variance formula (\ref{eqn:VarCij}) to a variety of checks for the lazy open quantum walk.

\subsection{The microscopic derivation}

Any CPTP map, such as the open quantum walk map described in Section \ref{sec:1}, may be thought of as a quantum channel. Given a quantum channel, the Stinespring dilation theorem \cite{Stinespring} guarantees the existence of a physical system implementing the given map. Thus, one may ask what is a physical system giving rise to the OQW? The first few steps in this direction were undertaken in \cite{2014IJQI1261010S} culminating in \cite{PhysRevA.92.032105}. The Hamiltonian for the total system may be written as the sum of the system, bath and system-bath interactions Hamiltonians,
\begin{eqnarray}
	H = H_{S} + H_{B} + H_{SB}.
	\label{eq:TotalHam}
\end{eqnarray}
The system Hamiltonian describes the local free evolution of the walker's internal degree of freedom as well as the position on the underlying graph. Thus, 
\begin{eqnarray}
	H_{S} = \sum\limits_{i}\Omega_{i}  \otimes \ket{i}\!\!\bra{i}.
	\label{eq:HamHS}
\end{eqnarray}
Concretely, the bath is thought of as a bath of harmonic oscillators with $H_{B}$ expressed in terms of bosonic creation and annihilation operators
\begin{eqnarray}
	H_{B} = \sum\limits_{i\neq j}\sum\limits_{n}\omega_{i,j,n} a^{\dagger}_{i,j,n}a_{i,j,n}.
	\label{eq:HamHB}
\end{eqnarray}
The system-bath interaction describs the bath driven transitions from site to site on the graph and hence may be written as
\begin{eqnarray}
	H_{SB} = \sum\limits_{i\neq j}\sum\limits_{n} Q_{i,j} \otimes X_{i,j} \otimes B_{i,j}.
	\label{eq:HamHSB}
\end{eqnarray}
The $Q_{i,j} \in \mathcal{B}\left(\mathcal{H}\right)$ operators are responsible for transforming the internal degree of freedom when a transition involving sites $i$ and $j$ occurs. The $X_{i,j} \in \mathcal{B}\left(\mathcal{K}\right)$ is responsible for implementing the steps between the sites. A simple Hermitian choice for $X_{i,j}$ is $X_{i,j} =  \ket{i}\!\!\bra{j} +  \ket{j}\!\!\bra{i}$. Lastly, the $B_{i,j} = \sum_{n}(g_{i,j,n}a_{i,j,n}+ g^{*}_{i,j,n}a^{\dagger}_{i,j,n})$ describes the coupling of the walker with the local environment. 

The microscopic derivation of the open quantum walk model, performed in \cite{PhysRevA.92.032105} for a graph with a general topology, employed the theory outlined in \cite{breuer2007theory}. Using the Born-Markov approximation the reduced density matrix of the system $\rho_{s}\left(t\right)$, in the interaction picture, satisfies the equation
\begin{eqnarray}
	&&\frac{d}{dt}\rho_{s}\left(t\right) = \\
	&&\hspace{25pt} -\int^{\infty}_{0} d\tau \mathrm{Tr}_{B}\big[ H_{SB}\left(t\right), \left[ H_{SB}\left(t-\tau\right), \rho_{s}\left(t\right) \otimes \rho_{B} \right] \big]\nonumber
	\label{eq:reduced}
\end{eqnarray}
where $\mathrm{Tr}_{B}$ stands for tracing out the bath degrees of freedom, and $\rho_{B}$ denotes the density matrix of the bath. Assuming that the environment is in a thermal equilibrium state, $\rho_{B} = \mathrm{exp}(-\beta H_{B})/\mathrm{Tr}[\mathrm{exp}(-\beta H_{B})]$. We assume that each of the $\Omega_{i}$'s have a unique set of eigenvalues. Their spectral decomposition may be written in terms of their eigenvalues $\lambda^{\left(i\right)}$ and orthogonal projectors $\Pi_{i}\left( \lambda^{\left(i\right)} \right)$. The $Q_{i,j}$ operators are then expressed in the basis associated with $\Omega_{i}$ and $\Omega_{j}$,
\begin{eqnarray}
	Q_{i,j}\left( \omega \right) &=& \sum\limits_{ \lambda^{\left(i\right)} - \lambda^{\left(j\right) } =\, \omega < 0} \Pi_{i}\left( \lambda^{\left(i\right)} \right)Q_{i,j}\Pi_{j}\left( \lambda^{\left(j\right)} \right),\\
	\label{eq:QijDecomp}
	Q^{\dagger}_{i,j}\left(\omega'\right) &=& Q_{i,j}\left( - \omega' \right).
	\label{eq:QdagijDecomp}
\end{eqnarray}
After transforming $H_{SB}$ to the interacting picture, and using the rotating wave approximation for the transition frequencies $\omega$ and $\omega'$, the following form for the master equation for $\rho_{s}\left(t\right)$ emerges
\begin{eqnarray}
	\frac{d}{dt}\rho_{s}\left(t\right) &=& \sum\limits_{i,j}\sum\limits_{\omega} \Big\{ \gamma_{i,j}\left( -\omega \right)\mathcal{D}\left[ Q_{i,j}\left(\omega\right) \otimes \ket{j}\!\!\bra{i} \right]\rho_{s}\left(t\right) \nonumber \\
	&& + \gamma_{i,j}\left( \omega \right)\mathcal{D}\left[ Q^{\dagger}_{i,j}\left(\omega\right) \otimes \ket{i}\!\!\bra{j} \right]\rho_{s}\left(t\right) \Big\}\nonumber\\
	&&\hspace{-25pt}+ \sum\limits_{i,j}\sum\limits_{\omega} \Big\{ \gamma_{i,j}\left( -\omega' \right)\mathcal{D}\left[ Q_{i,j}\left(\omega'\right) \otimes \ket{i}\!\!\bra{j} \right]\rho_{s}\left(t\right)\nonumber \\
	&& + \gamma_{i,j}\left( \omega' \right)\mathcal{D}\left[ Q^{\dagger}_{i,j}\left(\omega'\right) \otimes \ket{j}\!\!\bra{i} \right]\rho_{s}\left(t\right) \Big\}\nonumber\\
	\label{eq:MasterEqrhos}
\end{eqnarray}
where $\mathcal{D}\left(X\right)\rho$ denotes standard dissipative superoperator in Gorini-Kossakowski-Sudarshan-Lindblad (GKSL) form \cite{breuer2007theory, doi:10.1063/1.522979,Lindblad1976}
\begin{eqnarray}
	\mathcal{D}\left(X\right)\rho = X\rho X^{\dagger} - \frac{1}{2}X^{\dagger}X\rho - \frac{1}{2}\rho X^{\dagger}X.
	\label{eq:DXonrho}
\end{eqnarray}
In (\ref{eq:DXonrho}) the $X$'s form a basis for the corresponding $N$-dimensional Liouville space \cite{breuer2007theory}. The function $\gamma\left( \omega \right)$ is the real part of the Fourier transformation of the bath correlation functions $\langle B^{\dagger}_{i,j}\left(s\right)B_{i,j}\left(0\right) \rangle$. See \cite{PhysRevA.92.032105} for the full expression. After writing $\rho_{s}(t) = \sum_{i}\rho_{i}(t)\otimes \ket{i}\!\!\bra{i}$, one may derive a system of master equations
\begin{widetext}
\begin{eqnarray}
	&&\hspace{-60pt} \frac{d}{dt}\rho_{i}\left(t\right) = \sum\limits_{j,\omega}\gamma_{j,i}\left(-\omega\right) Q_{j,i}\left(\omega\right)\rho_{j}Q^{\dagger}_{j,i}\left(\omega\right) - \frac{\gamma_{i,j}\left(-\omega\right)}{2}\{ Q^{\dagger}_{i,j}\left(\omega\right)Q_{i,j}\left(\omega\right),\rho_{i} \}\nonumber\\
	&+&  \sum\limits_{j,\omega}\gamma_{i,j}\left(\omega\right) Q^{\dagger}_{i,j}\left(\omega\right)\rho_{j}Q_{i,j}\left(\omega\right) - \frac{\gamma_{j,i}\left(\omega\right)}{2}\{ Q_{j,i}\left(\omega\right)Q^{\dagger}_{j,i}\left(\omega\right),\rho_{i} \}\nonumber\\
	&+& \sum\limits_{j,\omega}\gamma_{i,j}\left(-\omega'\right) Q_{i,j}\left(\omega'\right)\rho_{j}Q^{\dagger}_{i,j}\left(\omega'\right) - \frac{\gamma_{j,i}\left(-\omega'\right)}{2}\{ Q^{\dagger}_{j,i}\left(\omega'\right)Q_{j,i}\left(\omega'\right),\rho_{i} \}\nonumber\\
	&+&  \sum\limits_{j,\omega}\gamma_{j,i}\left(\omega'\right) Q^{\dagger}_{j,i}\left(\omega'\right)\rho_{j}Q_{j,i}\left(\omega'\right) - \frac{\gamma_{i,j}\left(\omega'\right)}{2}\{ Q_{i,j}\left(\omega'\right)Q^{\dagger}_{i,j}\left(\omega\right),\rho_{i} \}.
	\label{eq:MasterEqrhosi}
\end{eqnarray}
\end{widetext}
This defines the continuous time OQW. To obtain the discrete time OQW of section \ref{sec:1}, a time step $\Delta$ is introduced and the time derivative in the differential equation is discretised in terms of $\Delta$. The connection between the discretised version of (\ref{eq:MasterEqrhosi}) and the discrete time OQW is established by the following identifications 
\begin{widetext}
\begin{eqnarray}
	&&\hspace{-70pt}B^{i\left(1\right)}_{j}\left(\omega\right) = \sqrt{\Delta \gamma_{j,i}\left(-\omega\right)} Q_{j,i}\left(\omega\right), \hspace{20pt}  
	B^{i\left(2\right)}_{j}\left(\omega\right) = \sqrt{\Delta \gamma_{i,j}\left(\omega\right)} Q^{\dagger}_{i,j}\left(\omega\right)\nonumber\\
	&&\hspace{-70pt}B^{i\left(1\right)}_{j}\left(\omega'\right) = \sqrt{\Delta \gamma_{i,j}\left(-\omega'\right)} Q_{i,j}\left(\omega'\right), \hspace{20pt}  B^{i\left(2\right)}_{j}\left(\omega'\right) = \sqrt{\Delta \gamma_{j,i}\left(\omega'\right)} Q^{\dagger}_{j,i}\left(\omega'\right)\nonumber\\
	&&\hspace{-70pt}B^{i}_{i}\left(\omega\right) = I_{N} - \frac{\Delta}{2}\sum\limits_{j,\omega}\Big( \gamma_{i,j}\left(-\omega\right)Q^{\dagger}_{i,j}\left(\omega\right)Q_{i,j}\left(\omega\right) + \gamma_{j,i}\left(\omega\right)Q_{j,i}\left(\omega\right)Q^{\dagger}_{j,i}\left(\omega\right) \Big)\nonumber\\
	&&\hspace{-40pt} - \frac{\Delta}{2}\sum\limits_{j,\omega'}\Big( \gamma_{i,j}\left(-\omega'\right)Q^{\dagger}_{j,i}\left(\omega'\right)Q_{j,i}\left(\omega'\right) + \gamma_{i,j}\left(\omega'\right)Q_{i,j}\left(\omega'\right)Q^{\dagger}_{i,j}\left(\omega'\right) \Big).
\label{eq:Identification}
\end{eqnarray}
\end{widetext}
One may now show that the OQW with the transition operators in equation (\ref{eq:Identification}) satisfies the normalization condition of (\ref{eq:KrausRep0}) up to $O(\Delta^{2})$, and the iteration formula for $\rho^{\left[n+1\right]}_{i}$, at time $n+1$, is of the same form as (\ref{eq:diagonaln}).
As one can see from the presence of the $B^{i}_{i}$ transition operator in (\ref{eq:Identification}), all microscopically derived OQWs are lazy. The expressions for the transition operators $B^{i}_{j}$ in  (\ref{eq:Identification}), and as described in \cite{PhysRevA.92.032105}, establish connections between the dynamical properties of the OQW and the thermodynamic properties of the environment.

In the remainder of this section, we specialise the microscopic derivation to a homogeneous discrete time OQW on the lattice $\mathbb{R}^{d}$, which is necessary for our central limit theorem to be applicable. We assume that the local unitary Hamiltonians on each site are identical and are denoted by $H_{0}$. Thus, for all $i$, $\Omega_{i} = H_{0}$. Next, recall for the homogeneous OQW map of the lattice, all the operators transforming the walker's internal degrees of freedom along a given axis, specifying a given direction, are identical. Thus, these operators were expressed in terms of the $A_{i}$'s and the $A_{i+d}$'s. In the microscopic derivation we now similarly have $Q_{i}$ and $Q_{i+d}$. Define the relation
\begin{eqnarray}
	A_{i} &=& \sqrt{\Delta}Q_{i}, \hspace{20pt} A_{i+d} = \sqrt{\Delta}Q_{i+d}, \nonumber \\
	A_{0} &=& I - \frac{\Delta}{2}\sum\limits^{2d}_{i=1}Q^{\dagger}_{i}Q_{i} - iH_{0}\Delta,
\label{eq:MicroDisc}
\end{eqnarray}
which corresponds to equations (\ref{eq:Identification}) for example at zero temperature. Using this definition we obtain the discrete time homogeneous OQW on the lattice $\mathbb{Z}^{d}$ from a microscopic derivation. In (\ref{eq:MicroDisc}), we have absorbed the $\gamma$ functions into the definition of the $Q$'s. Up to $O\left(\Delta^{2}\right)$, we have
\begin{eqnarray}
	\sum\limits^{2d}_{i=0} A^{\dagger}_{i}A_{i} = I.
\end{eqnarray}
The CPTP map on $\mathcal{H}$, after $n$ iterations, is
\begin{eqnarray}
	\tau^{\left(n+1\right)}_{i} = \sum\limits^{2d}_{j=0} A_{j}\tau^{\left(n\right)}_{i}A^{\dagger}_{j}.
\end{eqnarray}
By substitution of (\ref{eq:MicroDisc}) into the steady state condition $\rho_{\infty} = \mathcal{L}\left(\rho_{\infty}\right)$, it is straightforward to see 
\begin{eqnarray}
	&& 0 
	= - i  \left[ H_{0} , \rho_{\infty} \right] \nonumber\\
	&&+ \sum\limits^{2d}_{j=1} \left( Q_{j}\rho_{\infty}Q^{\dagger}_{j} - \frac{1}{2}Q^{\dagger}_{j}Q_{j} \rho_{\infty} -  \frac{1}{2} \rho_{\infty}Q^{\dagger}_{j}Q_{j} \right)
\label{eq:rhoGKSL}
\end{eqnarray}
up to $O(\Delta^{2})$. Note that the right-hand-side of equation (\ref{eq:rhoGKSL}) is precisely of GKSL form. The Lindbladian super-operator describes the time evolution of an open quantum system, with state $\rho$, and is defined by
\begin{eqnarray}
	\dot{\rho}
	=  - i  \left[ H , \rho \right] + \sum\limits^{N^{2}-1}_{j=1} \mathcal{D}\left(X_{j}\right)\rho.
\label{eq:Lindbladian}
\end{eqnarray}
Since the steady state is time-independent (and thus its time derivative vanishes), we obtain the quantum master equation for $\rho_{\infty}$ in GKSL form
\begin{eqnarray}
	 \dot{\rho}_{\infty} = L_{Lin}\left( \rho_{\infty} \right).
\label{eq:rhoGKSL2}
\end{eqnarray}
We note that equation (\ref{eq:rhoGKSL2}) is independent of the time step $\Delta$. The mean, once written in terms of the $Q$ operators is
\begin{eqnarray}
m &=& \Delta \sum\limits^{2d}_{j=0} \Tr\left(Q_{j}\rho_{\infty}Q^{\dagger}_{j} \right)e_{j}= \Delta \sum\limits^{d}_{j=1} \Tr\left( \widetilde{Q}_{j} \rho_{\infty} \right)e_{j},\nonumber\\
 \end{eqnarray}
where we have defined $\widetilde{Q}_{j} = Q^{\dagger}_{j}Q_{j} - Q^{\dagger}_{j+d}Q_{j+d}$.

Next, we study equation (\ref{eq:TheLEq}). The left-hand-side becomes
\begin{eqnarray}
	 &&L_{l} - \mathcal{L}^{\dagger}\left( L_{l} \right) = -\Delta \Big[  i\left[ H_{0} , L_{l} \right] \nonumber \\
	 && + \sum\limits^{2d}_{j=1}\left( Q^{\dagger}_{j}L_{l}Q_{j} -\frac{1}{2} L_{l}Q^{\dagger}_{j}Q_{j} -\frac{1}{2} Q^{\dagger}_{j}Q_{j}L_{l}   \right) \Big],
\end{eqnarray}
where the terms in braces define the adjoint of the Lindbladian, $L^{\dagger}_{in}$ \cite{breuer2007theory}. The right-hand-side of (\ref{eq:TheLEq}) becomes
\begin{eqnarray}
	&&\sum\limits^{2d}_{i=0}A^{\dagger}_{i}A_{i}\left( e_{i} \cdot l \right) - \left( m \cdot l \right)I \nonumber \\
	&& \hspace{20pt} = \Delta \sum\limits^{d}_{j=1}\left[ \widetilde{K}_{j} - \Tr\left( \widetilde{K}_{j}\rho_{\infty} \right)I \right] \left(e_{j} \cdot l \right).
\end{eqnarray}
Thus, equation (\ref{eq:TheLEq}) describes the time evolution of the $L_{l}$ operator in the Heisenberg picture
\begin{eqnarray}
 	\dot{L_{l}} = L^{\dagger}_{Lin}\left( L_{l} \right) &=& \sum\limits^{d}_{j=1}\left[ \widetilde{K}_{j} - \Tr\left( \widetilde{K}_{j}\rho_{\infty} \right)I \right] \left(e_{j} \cdot l \right).
\end{eqnarray}
This equation is also independent of the time step size $\Delta$.

\subsection{Example of lazy OQW in $1D$}

\begin{figure}[htb]
\centering
    \includegraphics[width=.45\textwidth]{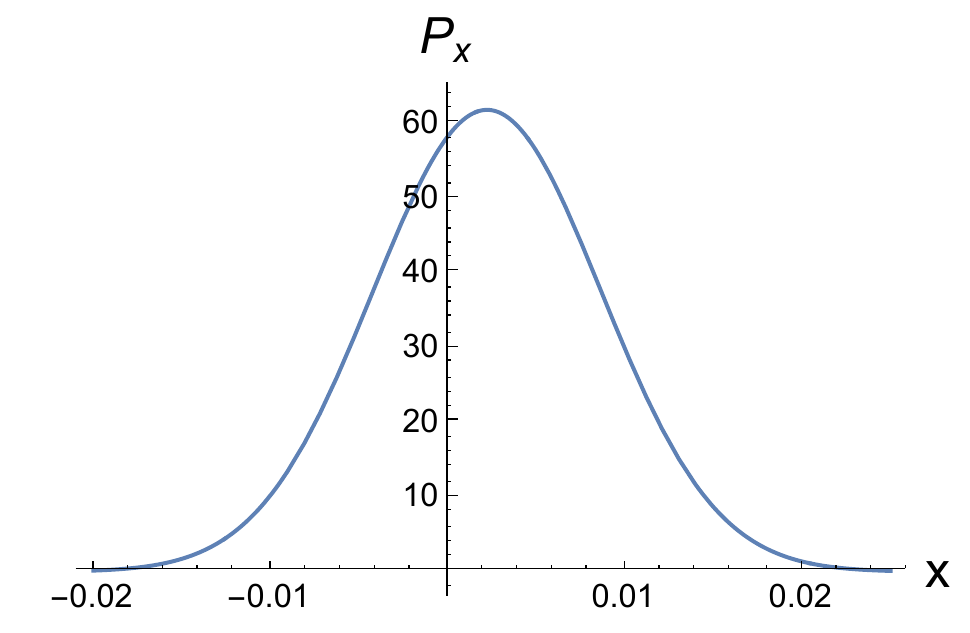}
  \caption{The Gaussian distribution plotted from the theoretically predicted values for the mean and variance. For $n_{y} = 1, \lambda = 0.3, \gamma = 0.1, \langle n \rangle = 1$ and $\Delta = 0.05$, we found $m = 0.00222$ and $C = 0.00645.$}
  \label{fig:Fig5}
\end{figure}

We turn now to some examples derived from the microscopic model. The first example, considered in \cite{PhysRevA.92.032105}, is the open quantum walk on the circle. The appropriate operators are
\begin{eqnarray}
	B &=& \sqrt{ \Delta \gamma\left( \langle n\rangle +1 \right)}\sigma_{-}, \hspace{20pt}C = \sqrt{ \Delta \gamma \langle n\rangle }\sigma_{+},\\
	A &=& I - \frac{\Delta}{2}\left[ \gamma \left( \langle n\rangle +1 \right)\sigma_{+}\sigma_{-} + \gamma \langle n\rangle \sigma_{-}\sigma_{+} \right]\\
	&& \hspace{90pt} - i \lambda \Delta \vec{n}_{\lambda} \vec{\sigma},\nonumber
\end{eqnarray}
where $ \vec{n}_{\lambda}  \vec{n}_{\lambda} = n^{2}_{x} + n^{2}_{y} + n^{2}_{z} = 1. $ Solving for the mean $m$ from $m = \mathrm{Tr}\left( B \rho_{\infty} B^{\dagger} \right) - \mathrm{Tr}\left( C \rho_{\infty} C^{\dagger} \right)$ we obtain
\begin{equation}
m = \Delta \frac{4\left( 1 - n^{2}_{z} \right)\gamma \lambda^{2}}{\gamma^{2}\left( 1+2\langle n\rangle \right)^{2} + 8\left( 1 + n^{2}_{z} \right)\lambda^{2} }.
 \label{eq:mean1dex2}
\end{equation}
Using formula (\ref{eqn:VarCij}) for the variance we find the following expression
\begin{widetext}
\begin{eqnarray}
	\sigma^{2} &=& \frac{4\Delta t_{-}\gamma\lambda^{2}}{s_{2}\left( s^{2}_{2}\gamma^{2} + 8 t_{+}\lambda^{2} \right)^{3}} \Big[ s^{6}_{2}\gamma^{4} + 8s^{2}_{2}t_{+}\gamma^{2}\lambda^{2}\left( 5n^{2}_{z} + 8 s_{1}\langle n\rangle -1 \right) + 64\lambda^{4}\left( s^{2}_{2} + 4n^{2}_{z}\left( s_{2} + 2\langle n\rangle \right) + n^{4}_{z}\left( 4s_{1}\langle n\rangle-1 \right) \right) \Big],\nonumber\\
\end{eqnarray}
\end{widetext}
where $t_{\pm} = 1 \pm  n^{2}_{z}$, and $s_{j} = 1 + j   \langle n\rangle$. An important check of formula (\ref{eqn:VarCij}) is that our expressions for the mean and variance agree with those in \cite{PhysRevA.92.032105}. Both $m$ and $\sigma^{2}$ indeed do reduce to the corresponding expressions in \cite{PhysRevA.92.032105} when $n_{y} = 1$. See Figure \ref{fig:Fig5} for the Gaussian plotted for this example. In this plot we chose the parameters $\lambda = 0.3$, $\gamma = 0.1$, $\langle n \rangle = 1$ and $\Delta = 0.05$. For these values, we find $m = 0.00222$ and $C = 0.00645$.

\subsection{$2D$ examples of lazy OQWs}

\subsubsection{Generic notations} In this subsection two examples of lazy OQWs in $2D$ will be presented. In both examples, the transition operators will follow from the outlined microscopic model for lazy OQWs in $2D$. To make notations more clear, the following conventions will be used:
\begin{itemize}
    \item coordinates on the $2D$ lattice $r=(i,j)$;
    \item possible movement from the $r$ along the $x$-axis is denoted $r_x=(i+1,j)$ and along the $y$-axis $r_y=(i,j+1)$, respectively;
    \item set of possible movements form the position $r$ is denoted as $r'=\{r_x,r_y\}$;\\
    for example
  \begin{eqnarray}
    	\sum_r f(r) &\equiv& \sum_{i,j} f(i,j) \hspace{10pt} or \nonumber \\
	\sum_{r,r'}f(r|r') & \equiv & \sum_r f(r|r_x)+f(r|r_y) \\
	&\equiv& \sum_{i,j}f(i,j|i+1,j)+f(i,j|i,j+1).\nonumber
    \end{eqnarray}
\end{itemize}

\subsubsection{Example 2}
Let us consider 2D array of two level atoms (for example, trapped ultra cold atoms on an optical lattice) described by the following Hamiltonian:
\begin{equation}
    H_S=\sum_{r}\frac{\omega_0}{2}\sigma_z\otimes|r\rangle\langle r|+\lambda\left(\vec{n}_\lambda\vec{\sigma}\right)\otimes|r\rangle\langle r|,
\end{equation}
where $\sigma_z$ is Pauli $z$ matrix and describes internal degree of freedom of the walker and $|r\rangle\equiv|i,j\rangle$ describes position of the on $2D$ lattice.
To end up with OQW on $2D$ one needs to assumes an environment assisted transport between every connected node of the walk.
Taking this into consideration the Hamiltonian of the bath reads,
\begin{eqnarray}
    H_B&=&\sum_{r,r'}\sum_n \omega_{r,r',n}a^\dagger_{r,r',n}a_{r,r',n},    
\end{eqnarray}
where operators $a^\dagger_{r,r',n}$ and $a_{r,r',n}$ denotes bosonic creation and annihilation operators of $n$-th mode of the thermal bath located between nodes $r$ and $r'$, the frequency of this mode is denoted by  $\omega_{r,r',n}$.

\begin{figure}
\centering
   \includegraphics[width=.45\textwidth]{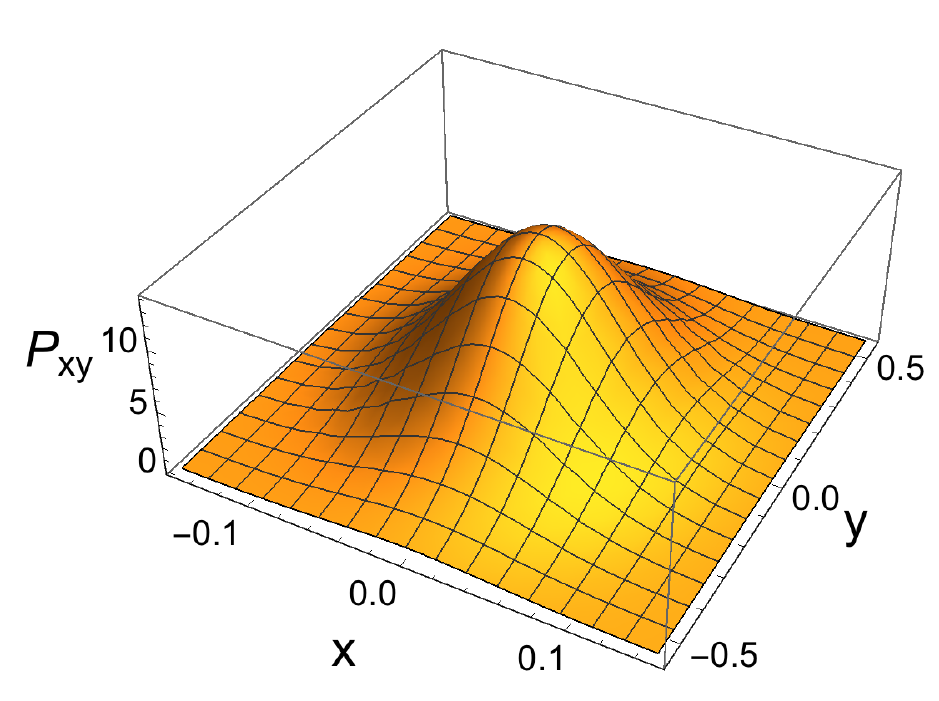}
  \caption{The Gaussian distribution plotted from the theoretically predicted values for the two-dimensional OQW in example 2. For $n_{y} = 1, \lambda = 0.3, \gamma = 0.1, \gamma^{+}_{y} = 0.5, \gamma^{-}_{y} = 0.5, \langle n \rangle = 1$ and $\Delta = 0.05$, we calculated $\vec{m}$ and $C$ given in equation (\ref{eq:Example2meanandCmatrix}). }
  \label{fig:Fig6}
\end{figure}

In this example, it is assumed that the transition of the walker along the $x$-axis is assisted via a dissipative coupling, while transition via $y$-axis is driven by the decoherent coupling. Under these assumptions the system-bath Hamiltonian $H_{SB}$ reads,
\begin{eqnarray}
    H_{SB} &=&\sum_{r,n} g_{r,r_x,n} a^\dag_{r,r_x,n} \sigma_-\otimes|r_x\rangle\langle r| +\mathrm{h.c.}\\\nonumber
    &+&\sum_{r,n} g_{r,r_y,n} a^\dag_{r,r_y,n} \sigma_z\otimes|r_y\rangle\langle r| +\mathrm{h.c.},
\end{eqnarray}
where coefficients $g_{r,r_i,n}$ denote the coupling strength between $n$-th mode of the bosonic bath located between nodes $r$ and $r_i$ with OQW walker. Following a generic microscopic derivation for OQWs \cite{PhysRevA.92.032105} it is straightforward to end up with the following transition operators,
\begin{eqnarray}
	B_{x} &=& \sqrt{ \Delta \gamma\left( \langle n\rangle +1 \right)}\sigma_{-}, \hspace{5pt}B_{y} = \sqrt{ \Delta \gamma^{+}_{y} \langle n\rangle }\sigma_{z},\\
	C_{x} &=& \sqrt{ \Delta \gamma\left( \langle n\rangle  \right)}\sigma_{+}, \hspace{5pt}C_{y} = \sqrt{ \Delta \gamma^{-}_{y} \langle n\rangle }\sigma_{z},\\
	A &=& I - \frac{\Delta}{2}\big[ \gamma \left( \langle n\rangle +1 \right)\sigma_{+}\sigma_{-} + \gamma \langle n\rangle \sigma_{-}\sigma_{+}  \nonumber
	\\&&\hspace{50pt}+ \gamma^{+}_{y}I + \gamma^{-}_{y}I\big] - i \lambda \Delta \vec{n}_{\lambda} \vec{\sigma}.
\end{eqnarray}
The mean in the $x$ and $y$ directions are
\begin{eqnarray}
	m_{x} &=& \frac{4 \gamma  \Delta  \lambda ^2 t_{-} T}{8 \lambda ^2 t_{-} T + \gamma  s_{2} \left(16 \lambda ^2 n_{z}^2+T^2\right)}\\
	m_{y} &=& \Delta r_{-},
\end{eqnarray}
where $r_{\pm} =  \gamma^{+}_{y}  \pm  \gamma^{-}_{y}$, $T = \gamma s_{2} + 4r_{+}$, $\eta^{\pm}_{z} = n^{2}_{z} \pm 1$. The covariance matrix entries are
\begin{widetext}
\begin{eqnarray}
C_{xx} &=& \frac{-4 \gamma  \Delta  \lambda ^2 T \eta^{-}_{z} }{{\left(\gamma  s_2 T^2+8 \lambda
   ^2 \left(\gamma  s_2 \eta^{+}_{z}-4 r_{+} \eta^{-}_{z}\right)\right)^3}}
   \Big[64 \lambda^4 \Big(\!\!-\!8 \gamma  r_{+} \eta^{-}_{z} \left(\left(8 \langle n\rangle
   s_1-s_2^2+1\right) n_z^2+s_2^2\right) \nonumber \\
   &&+\;\gamma^2 s_2 \left(s_2^2 \left(5 n_z^4-2
   n_z^2+1\right)-2 \left(8 \langle n\rangle s_1+3\right) n_z^2 \eta^{-}_{z}\right)+16
   r^2_{+} s_2 \left(\eta^{-}_{z}\right){}^2\Big) \nonumber \\
   &&+\;8 \gamma  \lambda ^2 \left(4 r_{+}+\gamma 
   s_2\right){}^2 \Big(\gamma  s_2 \left(\left(8 \langle n\rangle \left(2
   \langle n\rangle-s_1+2\right)+5\right) n_z^2+8 \langle n\rangle s_1-1\right) -4 r_{+} \left(8 \langle n\rangle
   s_1+1\right) \eta^{-}_{z}\Big) \nonumber \\&&+\gamma ^2 s_2^3 \left(4 r_{+}+\gamma 
   s_2\right)^4\Big]\\
	C_{yy} &=& \Delta r_{+}\\
	C_{xy} &=& C_{yx} = -\frac{16 \gamma ^2 \Delta  \lambda ^2 \eta^{-}_{z} r_{-} s_{2}
   \left(-16 \lambda ^2 n_{z}^2+16 r_{+}^2+8 \gamma  r_{+}+\gamma ^2
   s_{2}^2\right)}{\left(8 \lambda ^2 \left(2 \gamma  n_{z}^2
   s_{2}-n_{z}^2 T+T\right)+\gamma  s_{2} T^2\right)^2}.
\end{eqnarray}
\end{widetext}
Note that as $\gamma^{+}_{y} $ tends toward $\gamma^{-}_{y} $, the off diagonal entries $C_{xy}$ tend to zero. Figure \ref{fig:Fig6} shows the Gaussian distribution for this two-dimensional example. This Gaussian was plotted using the expressions derived for $\vec{m}$ and $C_{ij}$ for $n_{y} = 1, \lambda = 0.3, \gamma = 0.1, \gamma^{+}_{y} = 0.5, \gamma^{-}_{y} = 0.5, \langle n \rangle = 1$ and $\Delta = 0.05$. For these values of the parameters,
\begin{eqnarray}
   \vec{m} = \left(0.0009,0\right), \hspace{7pt} C = \left(\begin{array}{cc} 0.00261 & -0.00011 \\-0.00011 & 0.05 \end{array}\right).
  \label{eq:Example2meanandCmatrix}
  \end{eqnarray}

\subsubsection{Example 3}
\begin{figure}
\centering
   \includegraphics[width=.45\textwidth]{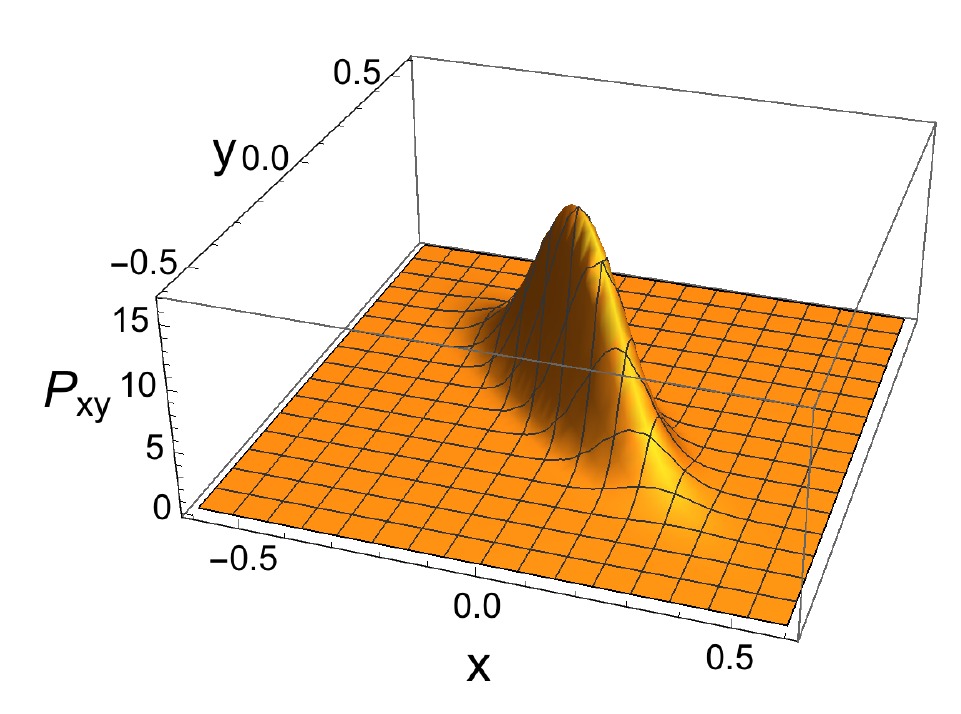}
  \caption{The Gaussian distribution plotted from the theoretically predicted values for the two-dimensional OQW in example 3. For $\lambda = 0.3, \gamma_{x} = 0.55, \gamma_{y} = 0.45, \langle n \rangle = 1$ and $\Delta = 0.05$, we found 
  the $\vec{m}$ and $C$ values given by equation (\ref{eq:Example3meanandCmatrix}).
   }
  \label{fig:Fig7}
\end{figure}

In this example, it is assumed that the transition of the walker along both axes is assisted via a dissipative coupling. From the microscopic point of view, this means that the only difference to the previous example would be in the form of the interaction Hamiltonian. Under these assumptions, the system-bath Hamiltonian $H_{SB}$ is as follows,
\begin{equation}
    H_{SB} =\sum_{r,r',n} g_{r,r',n} a^\dag_{r,r',n} \sigma_-\otimes|r'\rangle\langle r| +\mathrm{h.c.}.
\end{equation}
As in the previous example a generic microscopic derivation for OQWs \cite{PhysRevA.92.032105} would lead to the following form of the transition operators,
\begin{eqnarray}
	&&\hspace{-30pt}B_{x} = \sqrt{ \Delta \gamma_{x}\left( \langle n\rangle +1 \right)}\sigma_{-}, \hspace{2pt}B_{y} = \sqrt{ \Delta \gamma_{y}\left( \langle n\rangle +1\right) }\sigma_{-},\\
	&&\hspace{-30pt}C_{x} = \sqrt{ \Delta \gamma_{x} \langle n\rangle }\sigma_{+}, \hspace{10pt}C_{y} = \sqrt{ \Delta \gamma_{y} \langle n\rangle }\sigma_{+},\\
	&&\hspace{-30pt}A = I - \frac{\Delta}{2}\big[ \left(\gamma_{x}+\gamma_{y}\right) \left( \langle n\rangle +1 \right)\sigma_{+}\sigma_{-} \nonumber
	 \\ && \hspace{50pt} + \left(\gamma_{x} + \gamma_{y}\right) \langle n\rangle \sigma_{-}\sigma_{+} \big] - i \lambda \Delta \vec{n}_{\lambda} \vec{\sigma},
\end{eqnarray}
where $n_{y} = 1$ and $n_{x} = n_{z} = 0$. With these definitions we obtain the OQW for this particular model. The results are
\begin{eqnarray}
	m_{x} &=& \frac{4 \gamma_{x} \Delta  \lambda^2}{8 \lambda^2+(2  \langle n\rangle+1)^2
   (\gamma_{x}+\gamma_{y})^2},\\
    m_{y} &=& \frac{4 \gamma_{y} \Delta  \lambda^2}{8 \lambda^2+(2  \langle n\rangle+1)^2
   (\gamma_{x}+\gamma_{y})^2}.
\end{eqnarray}
The covariance matrix entries are
\begin{widetext}
\begin{eqnarray}	
C_{xx} &=& \frac{\Delta}{\left(8 \lambda ^2+r^2 s_2^2\right){}^2}\Big[-2 \langle n \rangle r s_2^2 \gamma _x^2 \left(4 \lambda ^2+(\langle n \rangle+1) r^2
   s_2\right) +\frac{8 \lambda ^2 r \gamma _x^2 \left(-4 \lambda ^2 (4 \langle n \rangle
   (\langle n \rangle+2)+3)-\langle n \rangle r^2 s_2^3\right)}{8 \lambda ^2+r^2 s_2^2}\nonumber   \label{eq:CxxEx3} \\
   &&+s_1 \gamma _x   \left(8 \lambda ^2+r^2 s_2^2\right) \left(4 \lambda ^2+\langle n \rangle r^2
   s_2\right) +\langle n \rangle \gamma _x \left(4 \lambda ^2+r^2 s_1 s_2\right) \left(8 \lambda^2+r^2 s_2^2\right)\Big]\\
   C_{xy} &=& -\frac{2 \Delta  \gamma _x \gamma _y \left(8 \lambda ^4 \left(8 \langle n \rangle^2+8
   \langle n \rangle+6\right) r s_2+\langle n \rangle r^5 s_1 s_2^5+16 \lambda ^2 \langle n \rangle r^3 s_1
   s_2^3\right)}{\left(8 \lambda ^2+r^2 s_2^2\right){}^3}.
  \label{eq:CxyEx3}
  \end{eqnarray}
  \end{widetext}
In (\ref{eq:CxxEx3}), $r = \gamma_{x} + \gamma_{y}$. $C_{yy}$ is the same expression but with $\gamma_{x}$ and $\gamma_{y}$ interchanged. We found the off-diagonal elements to be given by (\ref{eq:CxyEx3}). We note that $C_{yx} = C_{xy}$ and that the off-diagonal elements are symmetric under interchanging $\gamma_{x}$ and $\gamma_{y}$. Figure \ref{fig:Fig7} shows the Gaussian distribution for this final two-dimensional example. For the parameters, choosing $\lambda = 0.3, \gamma_{x} = 0.55, \gamma_{y} = 0.45, \langle n \rangle = 1$ and $\Delta = 0.05$
\begin{eqnarray}
    \vec{m} &=& \left(0.00102,0.00083\right), \\
     C &=& \left(\begin{array}{cc} 0.01829 & -0.01531 \\-0.01531 & 0.01775 \end{array}\right).
  \label{eq:Example3meanandCmatrix}
  \end{eqnarray}

\subsection{A numerical example}

In this section, we study a numerical example. We consider the matrices in equation (\ref{eq:ExampleOQW})
\begin{eqnarray}
A_{1} &=& \frac{1}{\sqrt{6}}\left(\begin{array}{cc}1 & 1 \\1 & e^{i\pi/3}\end{array}\right), \hspace{10pt} A_{0} =  \frac{1}{\sqrt{6}}\left(\begin{array}{cc}1 & e^{2i\pi/3} \\1 & -1\end{array}\right), \nonumber \\ A_{2} &=& \frac{1}{\sqrt{6}}\left(\begin{array}{cc}1 & e^{-2i\pi/3} \\1 & e^{-i\pi/3}\end{array}\right),
\label{eq:ExampleOQW1}
\end{eqnarray}
defining a lazy OQW on the line with a two-dimensional coin space. The steady state is
\begin{eqnarray}
	\rho_{\infty} = \left( \begin{array}{cc}
 0.5 & 0.375\, -0.217 i \\
 0.375\, +0.217 i & 0.5 \\
\end{array} \right),
\end{eqnarray}
leading to an $m$ value of zero. Equation (\ref{eq:TheLEq}), in vectorised form is (see equation (\ref{eq:vectorisation}) for the left-hand-side)
\begin{eqnarray}
&&\hspace{-20pt}\left(\begin{array}{cccc}
 0.5 & -0.5 & -0.5 & -0.5 \\
 0 & 1 & 0 & 0 \\
 0 & 0 & 1 & 0 \\
 -0.5 & -0.25-0.433 i & -0.25+0.433 i & 0.5 \\
\end{array}
\right)\left(
\begin{array}{c}
 L_{11} \\
 L_{12} \\
 L_{21} \\
 L_{22} \\
\end{array}
\right) \nonumber \\ &&\hspace{50pt}= \left(
\begin{array}{c}
 0 \\
 0.25\, +0.433 i \\
 0.25\, -0.433 i \\
 0 \\
\end{array}
\right)\!\!.
\label{eq:ExampleofLeq}
\end{eqnarray}
Solving this equation for the $L$ entries, we obtain
\begin{eqnarray}
	L = \left(
\begin{array}{cc}
 0.25 & 0.25\, +0.433 i \\
 0.25\, -0.433 i & -0.25 \\
\end{array}
\right).
\end{eqnarray}
Applying formula (\ref{eqn:VarCij}) to calculate the variance, we obtain $C = 1.04167$. To check this value, we simulated the above OQW for 10, 100, 1000, 10000 and 50000 steps. For each of these steps, we calculated the variance from the probability distribution and then converted it to the variance $C_{\mathrm{sim}}$ associated with the central limit theorem. Our results are summarized in Table \ref{Tab:1}.
\begin{table}[h!]
\begin{center}
\begin{tabular}{|c|c|}
	\hline
	$n$ & $C_{\mathrm{sim}}$  \\
	\hline
	10 & 1.08333
    \\
	\hline
	100 & 1.04583  \\
	\hline
	1000  & 1.04208 \\
	\hline
	10000 & 1.04171\\
	\hline 
	50000 & 1.04167 \\
	\hline
\end{tabular}
\end{center}
\caption{Table showing variance results from the OQW as defined in equation (\ref{eq:ExampleOQW}). The $n$ denotes the number of steps and the $C_{\mathrm{sim}}$ denotes the variance associated with the central limit theorem. The theoretical value for the variance was $C = 1.04167$. The table shows that the variance obtained from the simulation converges to the theoretical value as the number of steps $n$ increases.}
\label{Tab:1}
\end{table}

\section{Conclusion}

Many interesting and important results have been derived for open quantum walks in the non-lazy case. For example a central limit theorem for a homogeneous OQW on a $d$-dimensional lattice assuming a unique steady state. A microscopic derivation however revealed the necessity of formulating or extending the existing model to include a self jump. In this work we have provided such an extension. Establishing analogous results for the lazy OQW is non-trivial and would fill an important gap in the literature. We found that a central limit theorem exists for our lazy OQW. We obtained an analytic formula (\ref{eqn:VarCij}) for the variance of the associated Gaussian distribution. The quantities populating this expression, i.e. the steady state $\rho_{\infty}$, mean $m_{i}$ and the $L_{i}$ matrices, are found to be different from the formulation of the non-lazy OQW central limit theorem. We checked formula (\ref{eqn:VarCij}) for a number of examples. Three analytic examples were presented where the OQW was derived from specific microscopic models. A numerical example was then considered in which evidence was presented for the convergence of the variance, calculated from the simulated trajectories, to the variance calculated using (\ref{eqn:VarCij}). A key finding of this work shows that a central limit theorem may be applied to microscopically derived OQWs. 

Further insight was obtained into equation for the $L_{i}$ operators in equation (\ref{eq:TheLEq}). We found that the system is degenerate and that, up to a multiple of the identity matrix, the $L$ operators are Hermitian. For $L$ being a $D\times D$ matrix, this means that $\frac{1}{2}D\left(D-1\right)$ of the off-diagonal entries are not independent. We derived the discrete time homogeneous lazy OQW on the lattice $\mathbb{R}^{d}$ from the microscopic model. In terms of the operators from the microscopic model, we managed to write the time evolution for the steady state $\rho_{\infty}$ in GKSL form. Also in terms of the microscopic model operators, we wrote the time evolution for $L_{l}$ in the adjoint GKSL form.

One of the main assumptions in our work is that the OQW steady state is unique. The problem of formulating a central limit theorem for the case of a non-unique steady state is an interesting future avenue of research to pursue. One can indeed construct examples of OQWs that converge to multiple steady states. It is conceivable that a central limit theorem could potentially exist for each steady state, and that the corresponding analytic formulas would, in some way, depend on the initial state of the walk. 

In summary, we have defined a lazy open quantum walk in which the walker has the possibility of staying put on the same lattice site. We then derived a central limit theorem for our model on a homogeneous lattice and then presented evidence supporting our derivation. This work adds an important piece to the overall OQW framework.

\begin{acknowledgments}
This work was supported by the South African Research Chair Initiative of the Department of Science and Technology and the  National  Research  Foundation. IS acknowledge support in part by the National Research Foundation of South Africa (Grant No. 119345).
\end{acknowledgments}


\bibliography{Refs}

\end{document}